\documentclass[a4paper,11pt]{article}
\pdfoutput=1 % if your are submitting a pdflatex (i.e. if you have
\usepackage{jheppub} % for details on the use of the package, please
\usepackage[T1]{fontenc} % if needed
\usepackage{colortbl}
\usepackage{booktabs}
\usepackage{slashed}
\usepackage{xspace}
\usepackage{xcolor}
\usepackage{chngpage}
\usepackage{upquote}
\usepackage{multirow}
\usepackage{subcaption}
\usepackage{amssymb}

\usepackage{amsmath}

\usepackage{mmacells}
\mmaSet{
uselistings=false,
  morefv={gobble=2},
  linklocaluri=mma/symbol/definition:#1,
  morecellgraphics={yoffset=1.9ex}
}

\newcommand{\ii}{\mathrm{i}} %\ensuremath{\mathrm{i}}}
\newcommand{\Op}{\mathcal{O}}
\newcommand{\wc}{C}

\title{\boldmath How large can lepton mixing be?
}

\author[a]{J. de Blas,}
\author[a]{C. Giuliano,}
\author[b,c]{G. Guedes,}
\author[a]{R. Sánchez López,}
\author[a]{J. Santiago}

\preprint{}
\affiliation[a]{Departamento de Física Teórica y del Cosmos, Universidad de Granada,\\ Campus de Fuentenueva, E-18071 Granada, Spain}
\affiliation[b]{Deutsches Elektronen-Synchrotron DESY, Notkestr. 85, 22607 Hamburg, Germany}
\affiliation[c]{CERN, Theoretical Physics Department, Esplanade des Particules 1, Geneva 1211, Switzerland}
\emailAdd{deblasm@ugr.es}
\emailAdd{chiaragiuliano@ugr.es}
\emailAdd{guilherme.guedes@cern.ch}
\emailAdd{rafasl@ugr.es}
\emailAdd{jsantiago@ugr.es}

\abstract{
We show that, contrary to common expectations, the observed charged leptons can have a substantial mixing with new heavier fermions, at the level of 20$\%$. 
This can happen, in the language of effective theories, when the effect of mixing with heavier fermions vanishes at tree level in operators of mass-dimension 6 (or it is suppressed by the small charged lepton masses), a cancellation that can be naturally ensured by symmetries. 
Using a model that realizes this scenario we consider all current direct and indirect constraints and show that experimental constraints on the mixing are so mild that, given the current direct limit on the mass of the heavy fermions, theoretical considerations  
become the leading current constraints on the mixing. We also estimate the sensitivity to the mixing at future experiments, including the high-luminosity phase of the LHC and, most notably, the FCC-ee, and FCC-hh. We find a pattern in which the reach of direct searches in hadron machines makes theoretical considerations lead the limits while the precision of lepton machines can beat these theoretical considerations. We find that the FCC can finally reach \textit{per mille} precision in the mixing squared of the charged leptons.
}

\begin{document} 
\begin{flushright}
CERN-TH-2025-248
\end{flushright}
\maketitle
\flushbottom

\section{Introduction \label{sec:intro}}

The properties of the charged leptons have been measured to an excellent experimental precision, agreeing with the Standard Model (SM) predictions to a striking better than \textit{per mille} accuracy~\cite{ParticleDataGroup:2024cfk}.  The common lore dictates that the observed charged leptons cannot have a significant mixing with new fermions. However, this statement is implicitly based on the assumption that fermion mixing necessarily induces tree-level corrections to the gauge couplings of the SM fermions suppressed, at most, by two powers of the heavy fermion mass. In the language of effective field theories, this corresponds to a tree-level contribution to operators of mass dimension 6 of the generic form $\phi^\dagger \ii\overleftrightarrow{D}_\mu \phi \overline{\psi} \gamma^\mu \psi$. Indeed, this is true in the case of mixing with a single heavy fermion~\cite{delAguila:1982fs,delAguila:2000aa}, resulting in very stringent constraints on the allowed fermion mixing~\cite{deBlas:2013gla}. However, mixing with several different heavy fermions can induce cancellations that make this leading effects vanish\footnote{See~\cite{Chala:2025utt} for a recent study of models that generate other flat directions in electroweak (EW) precision data, even at the one-loop level.}. These cancellations can be ensured by a subset of custodial symmetry~\cite{Agashe:2006at}, common in realistic composite Higgs models~\cite{Anastasiou:2009rv}. The phenomenological implications at colliders of the corresponding sizeable mixing have been explored, for quarks in~\cite{Atre:2008iu,Atre:2011ae} and for tau leptons in~\cite{delAguila:2010es,Carmona:2013cq}. However, how large this mixing can be is a question that has not been properly addressed thus far.  The goal of the present work is to give a quantitative answer to this question, using all relevant current experimental data, as well as an estimation of the constraints that future colliders will put on this mixing.
In order to do this we consider a degenerate bi-doublet model that realizes the corresponding symmetry at tree level, ensuring the cancellation of the Wilson coefficients (WCs) of the dangerous dimension-6 operators at this order. The only remaining tree-level generated operator is suppressed by the charged lepton Yukawa couplings and, in practice, only modifies the lepton couplings to the Higgs boson. As we will see, this effect sets the most stringent, although still relatively mild, experimental bound on the allowed tau mixing while being irrelevant for the first two families.

In the absence of significant tree level, dimension 6 contributions, experimental constraints arise from tree-level generated dimension 8 operators or one-loop generated dimension 6 ones. In both cases the extra suppression results in milder bounds on the allowed mixing. The effective field theory (EFT) framework allows us to explore the indirect constraints from electroweak precision observables (EWPO), Higgs measurements, anomalous magnetic moments and flavor observables on this model. We also consider direct collider bounds on the model arising from pair- and single-production channels. 
The main result is that, with current data, the right-handed (RH) electron could have a non-singlet component as large as $20\%$ (in a sense that will be made explicit below) without any conflict with current experimental measurements. The corresponding numbers for the muon and tau are $18\%$ and $16\%$, respectively. These bounds are so mild that, when combined with direct searches for the new fermions, result in regions of the parameter space that are excluded on theoretical grounds, 
such as the requirement of the scalar potential stability. This condition strengthens the limits to $14\%$, $13\%$ and $16\%$ for mixing with the $e$, $\mu$, and $\tau$ leptons, respectively.
Despite the extra suppression in this model, flavor constraints do prevent a sizable simultaneous mixing of the vector-like leptons (VLLs) with two different SM charged leptons.

Measurements at future collider experiments will put more stringent limits on this mixing. By the end of the high-luminosity phase of the Large Hadron Collider (HL-LHC), the corresponding numbers will be $8\%$, $6\%$, and $5\%$ for mixing with $e$, $\mu$, and $\tau$, respectively, if we do not include theoretical considerations, going down to $5\%$ for the case of the electron and muon, and to $4\%$ for the case of the tau, if we include them.
The electron–positron Future Circular Collider (FCC-ee) will improve the precision of EWPOs and Higgs measurements to the point that indirect bounds will surpass the theoretical constraints. Under conservative (aggressive) assumptions for the theoretical uncertainties of the observables, the expected upper bounds decrease to $1.7\%$, $1.7\%$, and $0.7\%$ ($0.3\%$, $0.3\%$, and $0.4\%$) for mixing with $e$, $\mu$, and $\tau$, respectively. Theoretical considerations do not strenghten these bounds in this case.
Finally, the hadron–hadron Future Circular Collider (FCC-hh), with its significantly enhanced center of mass energy, will further tighten the direct searches constraints. In this regime, theoretical bounds once again dominate, and the projected upper limits are expected to reach roughly $0.2\%$ for all mixing cases.

While this paper performs a complete phenomenological analysis of a particular model, its conclusions should serve a broader purpose. First, it shows how symmetry considerations in the UV theory can have drastic consequences in the low-energy SMEFT interpretation, which are not captured if only one field extensions are considered; second, it shows the relevance of performing loop-level analysis, and how the extreme precision at the FCC-ee can even probe these symmetry-protected models.

The rest of this article is organized as follows. We describe the details of the model in Section~\ref{sec:model}. Current constraints from direct searches are discussed in Section~\ref{sec:directsearches}, while the indirect limits on the model are studied in Section~\ref{sec:indirect}. 
Having covered the different sources of experimental constraints, we discuss in Section~\ref{sec:theory_bounds} a series of bounds on the model arising from theoretical considerations.
Section~\ref{sec:combination} is devoted to the combination of all current constraints, and Section~\ref{sec:future_colliders} to the estimation of expected future limits from the HL-LHC, FCC-ee and FCC-hh. We conclude in Section~\ref{sec:conclusions}. We also include four appendices where we describe some of the technical details of our work.

\section{Model definition}
\label{sec:model}

Let us consider an extension of the SM that includes two VLLs in the following representations of the $SU(3)_\mathrm{C} \times SU(2)_\mathrm{L} \times U(1)_\mathrm{Y}$ gauge group (we follow the conventions in~\cite{delAguila:2008pw}),
\begin{equation}
    \Delta_1\sim(1,2,-1/2),\quad \Delta_3\sim(1,2,-3/2),
\end{equation}
with Lagrangian
\begin{equation}
    \mathcal{L}=\mathcal{L}_{\mathrm{SM}} 
    + \overline{\Delta}_1 [\ii \slashed{D} - M] \Delta_1
    + \overline{\Delta}_3 [\ii \slashed{D} - M] \Delta_3
    -\Big[ 
    \lambda_i^\prime (\overline{\Delta}_1  \phi
   +\overline{\Delta}_3  \tilde{\phi})
e_i    +\mathrm{h.c.}\Big],
\label{eq:lag_unbroken}
\end{equation}
where $e_i$ stand for the SM RH charged lepton singlets, with $i$ the corresponding flavor index, and we assume, without loss of generality, the charged lepton Yukawa couplings to be diagonal in the SM Lagrangian, $y^l_{ij}=y^l_i \delta_{ij}$. The crucial ingredient of the model is the fact that the mass, $M$, and the Yukawa couplings, $\lambda_i^\prime$, of both VLLs are identical. As mentioned in the introduction this can be naturally enforced by custodial symmetry.

After electroweak symmetry breaking (EWSB) the two heavy multiplets give rise to one state of electric charge $-2$, two states of charge $-1$, and one of charge $0$. Neglecting neutrino masses, the only relevant mixing occurs, after EWSB, among the charge $-1$ states. Their mass matrix reads
\begin{equation}
 \mathcal{M}_0=   \begin{pmatrix}
    m_i \delta_{ij} & 0 & 0 \\ 
    m_i^\prime & M & 0  \\
    m_i^\prime & 0 & M
     \end{pmatrix},
\end{equation}
where we have defined $m_i = y_i^l v$ and $m^\prime_i = \lambda^\prime_i v$, with $v\sim 174$ GeV the Higgs boson vacuum expectation value.
We can now perform a change of basis
\begin{equation}
\psi_i^{L,R}=(U_{L,R})_{ij} \psi_j^{\mathrm{phys}\, L,R},
\end{equation}
where $\psi_i$ represent the gauge eigenstates and $\psi_i^\mathrm{phys}$ the mass eigenstates. The unitary matrices $U_{L,R}$ diaginalize $\mathcal{M}_0$
\begin{equation}
U_L^\dagger \mathcal{M}_0 U_R=\mathcal{M}_D,
\end{equation}
with $\mathcal{M}_D$ the diagonal, physical mass matrix. In the following we give the result of this diagonalization up to order $v^2/M^2$
(see~\cite{Atre:2008iu} and \cite{delAguila:2010es} for the same model, assuming mixing with a single SM fermion, in the quark and lepton sectors, respectively, where exact expressions are given). 
We find
\begin{equation}
\mathcal{M}_D \approx \mathrm{diag}\left[m_i \left( 1- \left( \frac{m_i^\prime}{M}\right)^2 \right), M, M \left( 1 + \sum_i \left( \frac{m_i^\prime}{M}\right)^2 \right)  \right],
\end{equation}
where the $\approx$ symbol follows from neglecting higher order terms in $m_i/M$,
and the mixing matrices
\begin{equation}
U_L \approx 
\begin{pmatrix}
U_{L}^{(3\times 3)} & 0 & \sqrt{2} \frac{m_i m_i^\prime}{M^2} \\
-\frac{m_i m_i^\prime}{M^2} & - \frac{1}{\sqrt{2}} &  \frac{1}{\sqrt{2}} \\
-\frac{m_i m_i^\prime}{M^2} &  \frac{1}{\sqrt{2}} &  \frac{1}{\sqrt{2}} 
\end{pmatrix},
\end{equation}
and
\begin{equation}
U_R \approx 
\begin{pmatrix}
U_{R}^{(3\times 3)}(1-\frac{m^\prime_i m^\prime_j}{M^2}) & 0 & \sqrt{2} \frac{m_i^\prime}{M} \\
-\frac{m_i^\prime}{M} & - \frac{1}{\sqrt{2}} &  \frac{1}{\sqrt{2}} - \sum_i \frac{m_i^{\prime\,2}}{\sqrt{2}M^2}\\
-\frac{m_i^\prime}{M} &  \frac{1}{\sqrt{2}} &  \frac{1}{\sqrt{2}}- \sum_i \frac{m_i^{\prime\,2}}{\sqrt{2}M^2} 
\end{pmatrix},
\end{equation}
where 
\begin{equation}
U_{L,R}^{(3\times 3)}= 1_{3\times 3}+ \mathcal{O}(v^2/M^2),
\end{equation}
are 3 by 3 unitary matrices that are reabsorbed in the PMNS matrix. As can be understood from the above expressions, the largest mixing occurs for the RH fields, since the left-handed (LH) one is not only suppressed by an extra power of $v/M$ but also by a light Yukawa coupling. 
In this approximation, the mixing is given by
\begin{equation}
    \sin{\theta_i} \approx \sqrt{2} \frac{m^\prime_i}{M},
\end{equation}
with its square giving the percentage of non-singlet that the physical RH charged leptons have. 

In the physical basis the lepton couplings to the SM gauge bosons and the Higgs boson can be written, without loss of generality, as
\begin{align}
\mathcal{L}_Z=& \frac{g}{2c_W} \bar{\psi}^i_Q \gamma^\mu \Big[ (-1)^{Q}X^{QL}_{ij} P_L + (-1)^{Q}X^{QR}_{ij} P_R - 2 s_W^2 Q \delta_{ij}\Big]
\psi_Q^j Z_\mu, \label{eq:LZ}
\\
\mathcal{L}_W =& \frac{g}{\sqrt{2}} \bar{\psi}^i_Q \gamma^\mu \Big[V^{QL}_{ij} P_L+V^{QR}_{ij} P_R\Big] \psi^j_{(Q-1)} W^+_\mu + \mathrm{h.c.}, \label{eq:LW}
\\    
\mathcal{L}^H=&-\frac{1}{\sqrt{2}} H \Big[\bar{\psi}^i_Q Y^Q_{ij} P_R \psi^j_Q + \mathrm{h.c.}\Big], 
\label{eq:LH}
\end{align}
where $Q=-2,-1,0$ represents the electric charge of the different particles, $P_{LR}=(1\mp \gamma^5)/2$ are the chirality projectors and $i,j$ run over all the particles with the corresponding electric charge.
We have introduced an explicit sign in front of the definition of the  $X$ matrices to ensure that in the SM limit they reduce to the identity matrix. Apart from the SM-like leptons, the spectrum consists of three degenerate states, $N$, $E_1$ and $Y$, of mass $M$ and charges 0, -1 and -2, respectively,
\begin{equation}
    m_N=m_{E_1}=m_Y=M,
\end{equation}
and one heavier state, $E_2$, of charge -1 and mass
\begin{equation}
m_{E_2}\approx M
\left(1 + \sum_i \left(\frac{m^\prime_i}{M}\right)^2\right).
\end{equation}
The masses of the SM-like charged leptons are
\begin{equation}
m_i^\mathrm{phys}\approx m_i \left(1- \left(\frac{m^\prime_i}{M}\right)^2\right),    
\end{equation}
with $i=1,2,3$ standing for $e$, $\mu$ and $\tau$, respectively. For simplicity, we will assume here and in the next  section that $\lambda_i^\prime$ (and therefore $m_i^\prime$) are real.

Considering the following basis, $e, \mu, \tau, E_1, E_2$ for the charge -1 particles and $\nu_e, \nu_\mu, \nu_\tau, N$ for the charge 0 ones, the couplings are, 
\begin{align}
X^{0L}=& \,\, 1_{4\times4},
\quad 
X^{0R}=  \begin{pmatrix} 0_{3\times 3} & \vec{0}  \\
\vec{0}^T & 1   
\end{pmatrix}, 
\\
%%%%%%
X^{-1L}\approx & \begin{pmatrix} 1_{3\times 3} & 
\sqrt{2} \frac{m_i m_i^\prime}{M^2} &
\sqrt{2} \frac{m_i m_i^\prime}{M^2} \\
\sqrt{2} \frac{m_i m_i^\prime}{M^2} &
0 & -1 \\
\sqrt{2} \frac{m_i m_i^\prime}{M^2} &
-1 & 0 
\end{pmatrix},
\quad X^{-1R}\approx  \begin{pmatrix} 0_{3\times 3} & \sqrt{2} \frac{m^\prime_i}{M} & \vec{0}  \\
\sqrt{2} \frac{m^\prime_i}{M} & 0 & -1+ \sum_i \frac{m_i^{\prime\,2}}{M^2}   \\
\vec{0}^T  & -1+ \sum_i \frac{m_i^{\prime\,2}}{M^2}  & 0
\end{pmatrix},
\\
%%%%%%
X^{-2L}=& \,\, X^{-2R}=-1,
\\
%%%%%%
V^{0L}\approx & \begin{pmatrix}
\tilde{V}_{3\times 3} &  \vec{0} &  \sqrt{2}\frac{m_i m_i^\prime}{M^2}
\\
-\frac{m_i m_i^\prime}{M^2}
& -\frac{1}{\sqrt{2}}
& \frac{1}{\sqrt{2}}
\end{pmatrix},
\quad
\quad V^{0R}\approx  \begin{pmatrix}
0_{3\times 3} 
& \vec{0}
&\vec{0} 
\\
-\frac{m_i^\prime}{M}
 & -\frac{1}{\sqrt{2}}
& \frac{1}{\sqrt{2}} \left(1
-\sum_i \frac{m_i^{\prime \,2}}{M^2}\right)
\end{pmatrix},
%%%%%%
\\
V^{-1L}\approx & \begin{pmatrix}
-\frac{m_i m_i^\prime}{M^2} &\frac{1}{\sqrt{2}}&\frac{1}{\sqrt{2}}
\end{pmatrix}^T, 
\quad
V^{-1R}\approx  \begin{pmatrix}
-\frac{m_i^\prime}{M} &\frac{1}{\sqrt{2}}&\frac{1}{\sqrt{2}}\left(1
-\sum_i \frac{m_i^{\prime \,2}}{M^2}\right)
\end{pmatrix}^T, 
\\
v Y^{-1} \approx & \begin{pmatrix}
m_i^{\mathrm{phys}} \left[\delta_{ij} -2\frac{m_i^\prime m_j^\prime}{M^2} \right]& \vec{0} & \sqrt{2} \frac{m_i m_i^\prime}{M}
 \\
\vec{0}^T & 0 & 0\\
\sqrt{2}m_i^\prime  
\left[1+ \frac{1}{M^2} \left(
m_i^2-m_i^{\prime\,2}- 2m_i^2\sum_{j\neq i}\frac{2 m_j^{\prime\,2}}{m_i^2-m_j^2}
\right)
\right]
& 0& 
2 \sum_i \frac{m_i^{\prime\,2}}{M} 
\end{pmatrix}.
\end{align}

\begin{figure}[h!]
    \centering
    \includegraphics[width=0.7\linewidth]{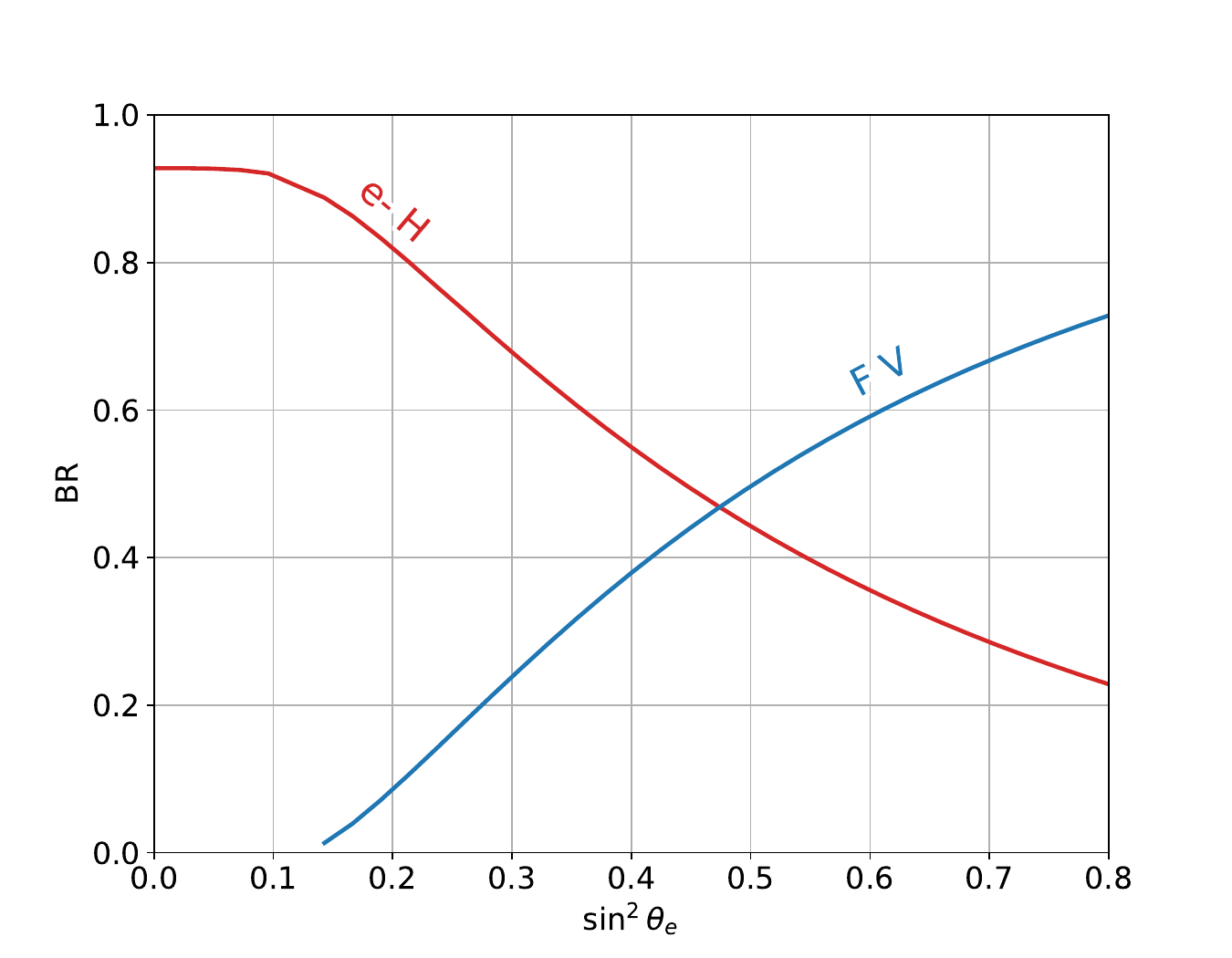}
    \caption{Leading branching ratios of the heaviest state, $E_2$, as a function of the mixing squared $\sin^2{\theta_e}$, at $M=1350$ GeV. The decay channels into a heavy state $F=E_1, N, Y$ and a gauge boson $V=Z, W^-,W^+$, all with the same branching fraction, are merged into the channel $E_2\to FV$.}
    \label{fig:BR_e2_1350}
\end{figure}

Note that the heaviest state, $E_2$, has order one couplings to the channels $Z E_1$, $W^+ Y$ and $W^- N$, and couplings proportional to $\lambda_i^\prime$ to the channels $e_i H$, with $e_i$ the three light, mostly SM-like, charged leptons. Thus, the branching ratios of $E_2$ are a function of $M$ and $m_i^\prime$ that we plot, assuming for simplicity mixing with a single SM charged lepton, in Figure~\ref{fig:BR_e2_1350}. As we see in the figure, $E_2$ decays predominantly into $e_i H$ for small values of the mixing, for which the decay into the states of mass $M$ is kinematically forbidden. Once the mixing increases, these decay channels become kinematically allowed and the decay into $Z E_1$, $W^+ Y$ and $W^- N$ start dominating (all with equal branching fraction). We have checked that the result is relatively independent of the value of $M$.
The remaining heavy states decay into SM particles, with branching fractions, at leading order in the mixing, equal to
\begin{equation}
BR(Y\to W^- e_i)\approx BR(E_1 \to Z e_i)\approx BR(N \to W^+ e_i)\approx \frac{m_i^{\prime\,2}}{\sum_j
m_j^{\prime\,2}}.   
\end{equation}

The expressions above allow us to consider the limits that direct searches impose on the masses and mixing of the heavy particles, as we will describe in the next section.

\section{Direct searches\label{sec:direct_searches}}
\label{sec:directsearches}

Due to the usual assumption that charged lepton mixing is very strongly constrained, direct searches for VLLs at colliders have typically focused on pair production, as this production mechanism is independent of the mixing angle. On the other hand, mixing could be larger for neutrinos and searches for heavy neutral leptons (HNLs) have also treated the single production as a discovery channel. In the following, we will consider both production modes, recasting existing analyses for the case of our model.

\subsection{Pair production}

Pair production of new VLLs is independent of the mixing, being sensitive only to the mass of the new particles. Direct searches of VLL are usually interpreted in terms of a charge -1 VLL singlet or a VLL doublet ($\mathrm{VLL^D}$ hereafter), that is a VLL with the same quantum numbers as the LH lepton doublet in the SM. Due to its proximity to our model, we will consider the latter interpretation to recast it in our case. 

\begin{figure}[t!]
    \centering
    \includegraphics[width=0.9\linewidth]{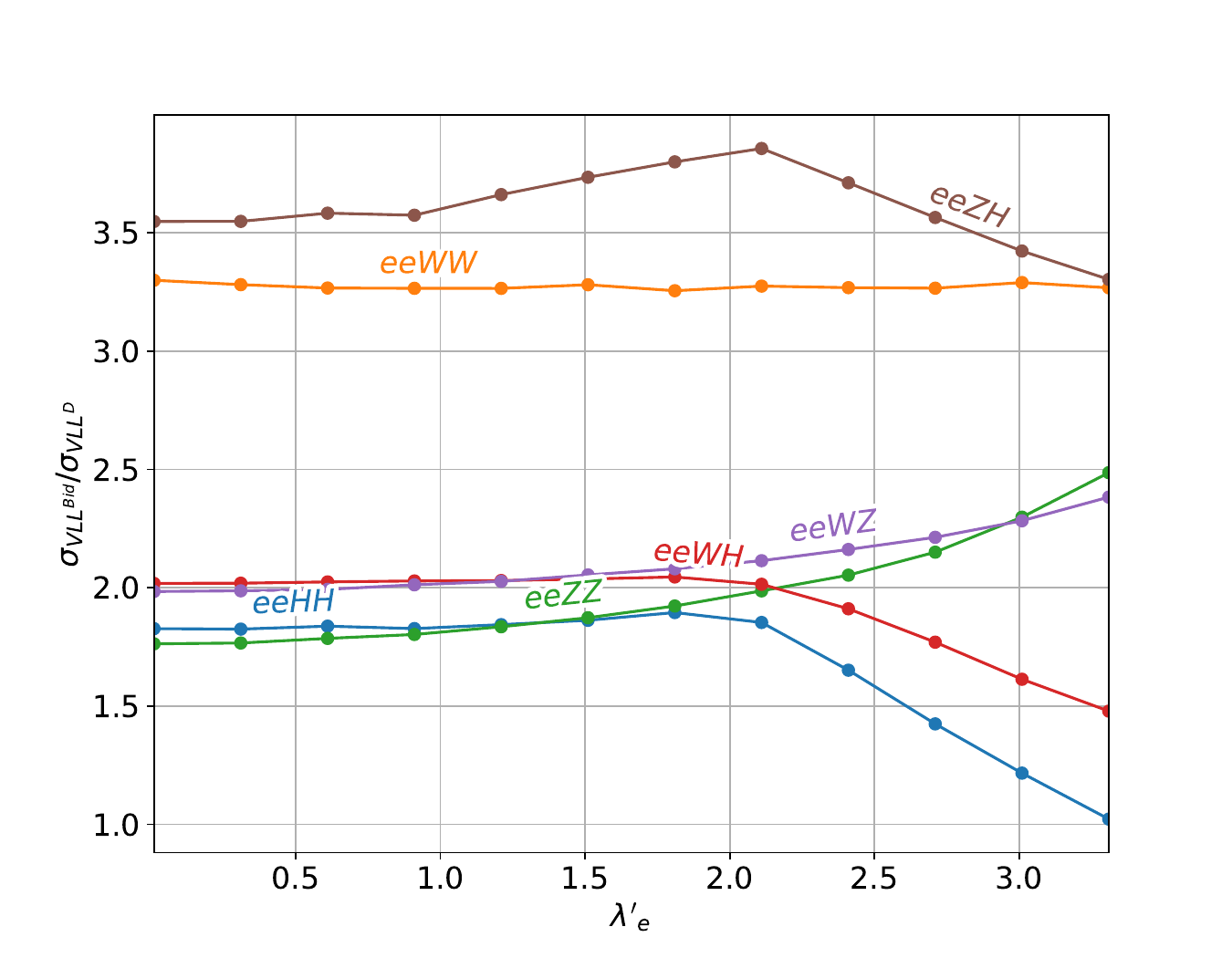}
    \caption{Ratio of the cross section in our model over the one in the $\mathrm{VLL^D}$ model, for the different decay channels, as a function of the coupling $\lambda^\prime_e$. We are considering the contribution of the heavy lepton pair production with decays in the relevant channels, for $M=1.35$ TeV.}
    \label{fig:xsec_comparison}
\end{figure}

Current searches for VLL have become very sophisticated (recent examples are~\cite{ATLAS:2023sbu,ATLAS:2024mrr} from ATLAS and \cite{CMS:2022nty} from CMS), including many different final states, that target all relevant production mechanisms and decay channels. All these channels are combined using machine learning tools, like deep neural networks or boosted decision trees, which makes the detailed recasting of the different analyses a significant challenge. Given the similarities between our model and the $\mathrm{VLL^D}$ we have numerically computed the ratio of the production cross section in all different channels for our model and for the $\mathrm{VLL^D}$ one. We have used \texttt{MG5\_NLO}~\cite{Alwall:2014hca, Frederix:2018nkq} to compute the corresponding cross section at tree level, with a result shown in Figure~\ref{fig:xsec_comparison}. We see that, generically, the cross section for the different channels is a factor of $\sim 2-3$ larger in our model than in the $\mathrm{VLL^D}$ one. This enhancement, which can be easily understood in terms of the couplings of the different states to the EW gauge bosons, is pretty independent of the mixing, except for large values of $\lambda^\prime$, for which the $E_2 \to F V$ decays, with $F=E_1, N, Y$ and $V= Z, W^-, W^+$, respectively, become kinematically open. In this case the decay channels including the Higgs boson get penalized and the ones involving extra vector bosons become enhanced.

Tailored searches, that take advantage of the particular features of our model, along the lines of the one in~\cite{Guedes:2021oqx}, will certainly be more restrictive than current ones, but we will defer such a study to future work. In this article we will consider an enhancement factor of 2.5 in the production cross section of our model with respect to the $\mathrm{VLL^D}$ one. This is the number we obtain assuming that all channels contribute equally. Using 2 or 3 for the enhancement factor results in a $\sim 3\%$ difference in the heavy mass limit which, in turn, implies a negligible difference in the limit on the mixing.

Using the latest ATLAS analysis~\cite{ATLAS:2024mrr}, and rescaling their results by a factor of 2.5, in the case of coupling to the first generation we obtain a constraint $M\geq 1390$ GeV. This number changes to $1350$ GeV and $1420$ GeV for an enhancement factor of 2 and 3, respectively. The same analysis can be carried out with VLL$_\mu^D$ \cite{ATLAS:2024mrr} and VLL$_\tau^D$ \cite{CMS:2022nty} model searches, where the subscript describes the generation to which the VLL couples to. Rescaling the exclusion plots by a factor 2.5 we obtain $M\geq 1430$ GeV and $M\geq 1210$ GeV for mixing with the $\mu$ and $\tau$, respectively. Again, using an enhancement factor of 2 or 3 results on a variation on the limit on the mass of the order of $3\%$. Thus, we take, as the final estimate of the current limit on $M$ from direct searches in pair production, the following numbers, at $95\%$ C.L.,
\begin{equation}
    M\geq \left \{
    \begin{array}{l}
    1390\mbox{ GeV, mixing with $e$}, \\
    1430\mbox{ GeV, mixing with $\mu$}, \\
    1210\mbox{ GeV, mixing with $\tau$}, 
    \end{array}
    \right.
    \mbox{   [direct searches, pair production].}
    \label{limit:direct:pair}
\end{equation}

\subsection{Single Production}

\begin{figure}[t!]
\centering
\begin{subfigure}{.4\textwidth}
  \centering
  \includegraphics[width=.8\linewidth]{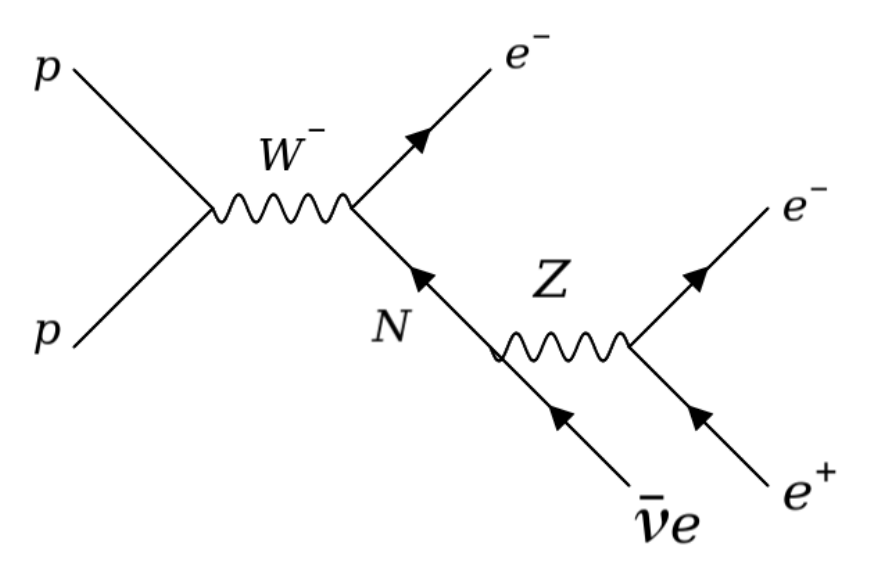}
  \caption{$eee$ signal through $N\to Z\nu_e$}
  \label{fig:zchannel}
\end{subfigure}
\begin{subfigure}{.4\textwidth}
  \centering
  \includegraphics[width=.8\linewidth]{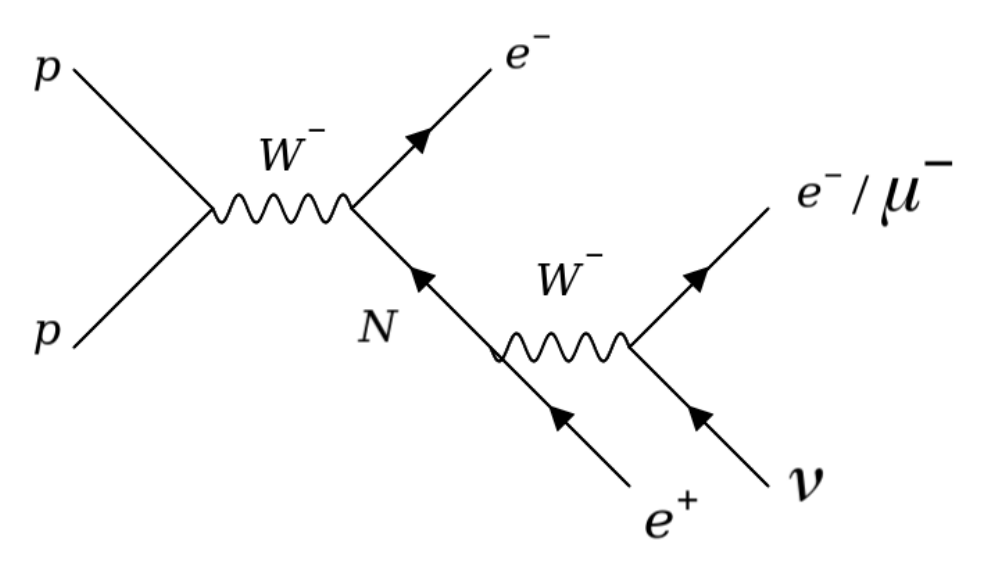}
  \caption{$eee/ee\mu$ signal through $N\to We$}
  \label{fig:wchannel}
\end{subfigure}
\\
\begin{subfigure}{.4\textwidth}
  \centering
  \includegraphics[width=.8\linewidth]{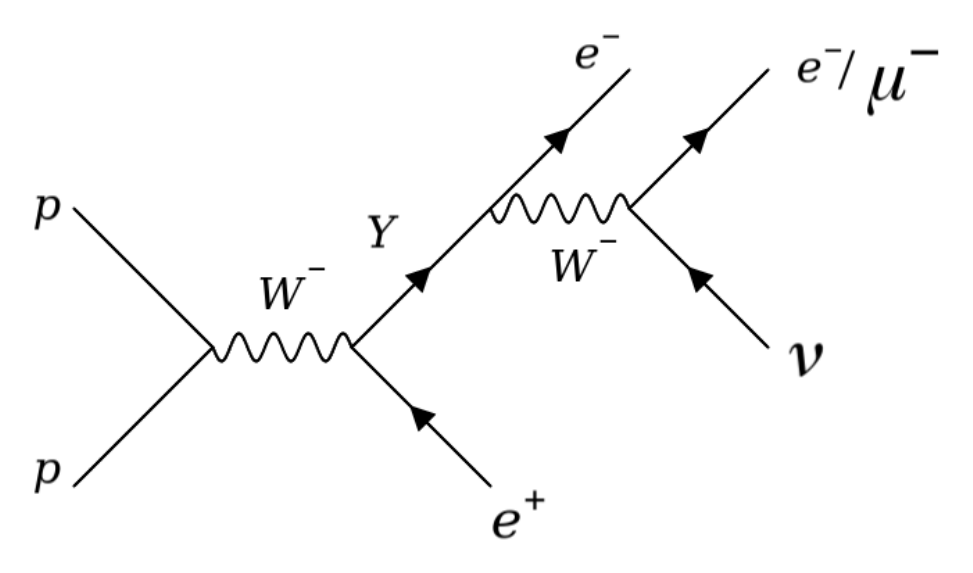}
  \caption{$eee/ee\mu$ signal through $Y\to We$}
  \label{fig:ychannel}
\end{subfigure}
\caption{$eee$ and $ee\mu$ channels via singly produced VLL. Diagram~\ref{fig:zchannel} is only present in the HNL model, diagram~\ref{fig:wchannel} is present (with different branching ratios) in both models and diagram~\ref{fig:ychannel} is only present in our model.}
\label{fig:HNLchannels}
\end{figure}
As mentioned above, single production of new heavy leptons has been considered so far only in the case of HNLs. Results are usually interpreted in terms of a single neutral VLL singlet, $N$. 
The results for Dirac HNL
can be easily reinterpreted in our model by identifying the relevant coupling to the $W$ boson, $V_{eN} \equiv m^\prime/M$.  
The main differences are the decay branching ratios and the fact that there is an extra state, $Y$, in our model that can be also singly produced with the same final state. The relevant diagrams are shown in Figure~\ref{fig:HNLchannels}. Diagram~\ref{fig:zchannel} is only present in the HNL case, with a $BR(N \to Z \nu)\approx 0.25$. Diagram~\ref{fig:wchannel} is present in both models, but with different branching ratios, $BR(N\to W e)\approx 0.5$ in the HNL case and $BR(N\to W e)=1$ in our model. Finally, diagram~\ref{fig:ychannel} is only present in our model, again with a 100$\%$ branching ratio. If the channels with a $W$-boson in the final state were the only ones present, we would expect a factor of 4 enhancement in the corresponding cross section (two channels with same production cross section and twice the decay ratio). In practice, there is also the channel with $Z$-boson in the final state, and therefore, the corresponding enhancement factor is a bit smaller than 4, with some mild dependence on the value of $M$.

\begin{figure}[h!]
    \centering
    \includegraphics[width=0.7\linewidth]{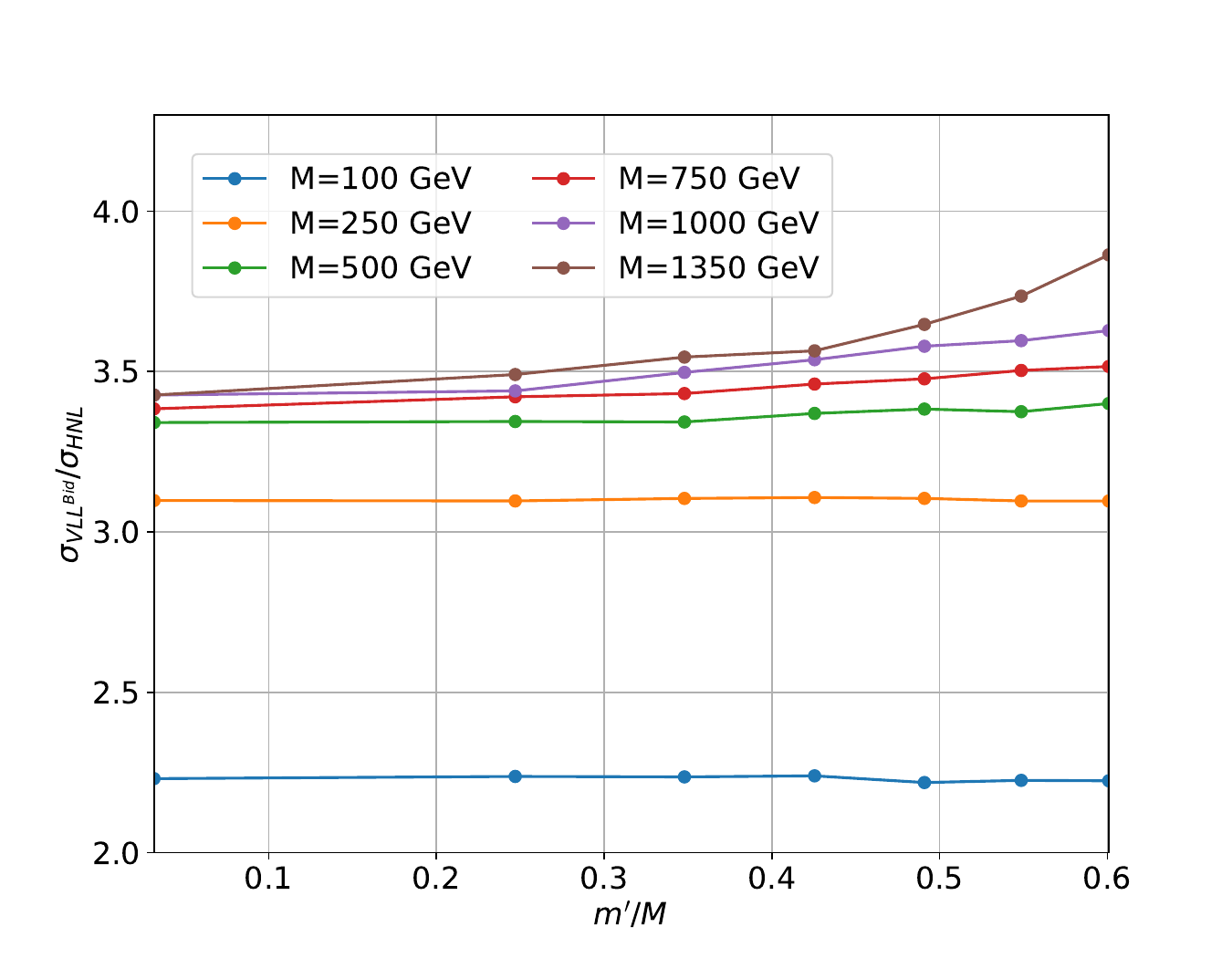}
    \caption{Ratio of the $3e$+missing energy production via heavy leptons in our model over the one in the HNL case, as a function of the corresponding coupling (recall the identification $V_{eN} \to m^\prime/M$) for different values of $M$.}
    \label{fig:2VLLevsHNL_def}
\end{figure}

\begin{figure}[t!]
\centering
  \includegraphics[width=.32\linewidth]{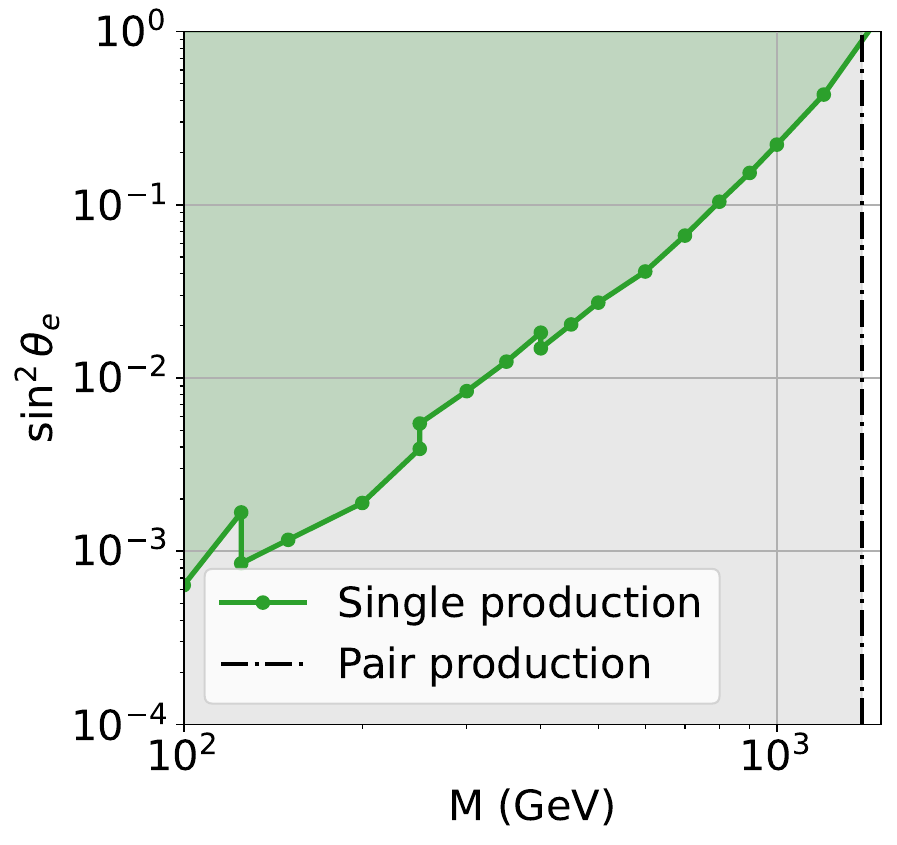}
  \includegraphics[width=.32\linewidth]{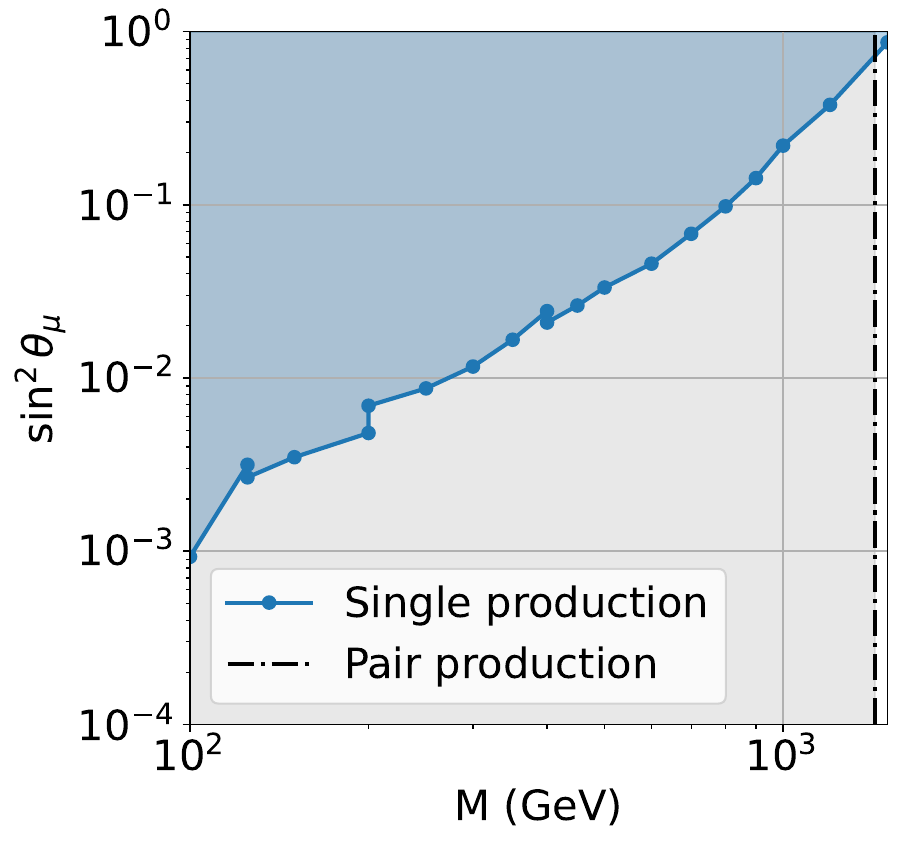}
 \includegraphics[width=.32\linewidth]{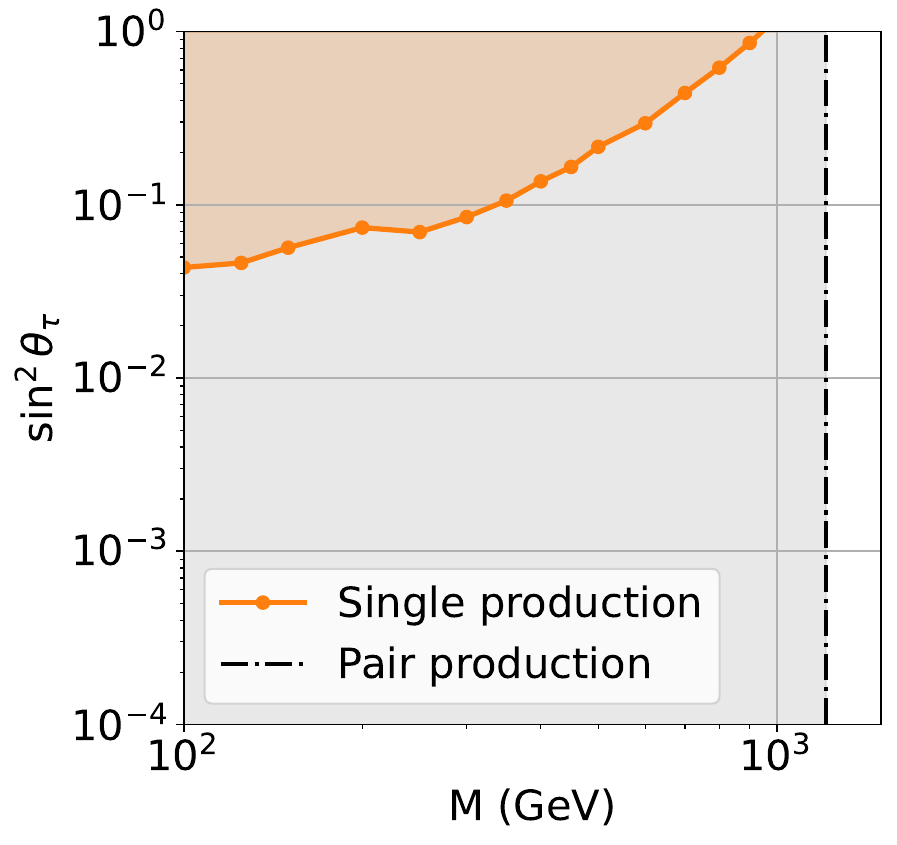}

\caption{ 95\% C.L. exclusion limits on $\sin^2{\theta}$ as a function of $M$ from direct searches in our model for the case of mixing with $e$ (left), $\mu$ (center) and $\tau$ (right). The excluded region from single production is shown as a colored shaded area, while the one from pair production is shown in grey. Only the white area is allowed by direct searches.}
\label{fig:2VLL_direct_searches}
\end{figure}

We have numerically computed the production of $3e$ plus missing energy via single production of heavy leptons both in the HNL case and in our model. The ratio of the two cross sections is shown in Figure~\ref{fig:2VLLevsHNL_def}, as a function of the corresponding coupling, for different values of $M$. As we can see in the figure, the ratio increases with $M$, converging to a value $\sim 3.5$ for $M\gtrsim 500$ GeV and it is insensitive to $m^\prime/M$ except for very large values of this parameter (and the mass $M$). 
We have used this enhancement factor for each value of $M$ to recast the results from~\cite{CMS:2024xdq} in our model. The corresponding limit on the mixing squared is shown in the left panel of Figure~\ref{fig:2VLL_direct_searches}, with the solid green line. 
We also show in the same plot the limit from pair production ($M\geq 1390$ GeV) with the vertical dash-dotted black line. It should be pointed out that this pair production limit exceeds the highest experimental point in the single production searches. In any case, as we will see in the next section, indirect constraints will put complementary constraints that make the ones from single production irrelevant.
The corresponding limits for mixing with the $\mu$ and $\tau$ leptons are shown in the central and right panel of Figure~\ref{fig:2VLL_direct_searches}, respectively.

\section{Indirect constraints\label{sec:indirect}}

Having computed the current limits on our model from direct searches, we turn our attention to indirect constraints. These can be most efficiently studied by integrating out from Eq.~(\ref{eq:lag_unbroken}) the heavy states, and matching the result onto the SM EFT (SMEFT). 
(Note that the stringent limit on their mass from direct pair production, Eq.~\eqref{limit:direct:pair}, combined with the fact that we are considering mostly low-energy observables ensures the validity of the EFT approach.)
We will do this in three different steps, starting with the tree-level contribution to dimension 6 operators in section~\ref{sec:tree-level-dim6} and to dimension 8 ones in section~\ref{sec:tree-level-dim8}, while section~\ref{sec:one-loop-dim6} will be devoted to the analysis of the one-loop contributions to the dimension 6 operators.

\subsection{Tree-level, dimension 6 effects \label{sec:tree-level-dim6}}

The degeneracy of the masses and couplings for the two new doublets ensures some cancellations in the tree-level contribution to the dimension 6 SMEFT Lagrangian. Indeed, in the Warsaw basis, the only non-vanishing contribution is to the operator
\begin{equation}
    \big(\mathcal{O}_{e\phi}\big)_{ij} = \phi^\dagger \phi \,\overline{l}_i \phi e_j,
\end{equation}
where $l_i$ and $e_j$ are the SM lepton doublet and singlet, respectively, with $i,j$ flavor indices. The corresponding WC reads~\cite{delAguila:2008pw,deBlas:2017xtg}
\begin{equation}
    (\wc_{e\phi})_{ij}=\frac{y^l_i (\lambda^\prime_i)^\ast \lambda^\prime_j} {M^2}.
\end{equation}

After the EWSB, this operator modifies the charged lepton masses and Yukawa couplings with the Higgs boson (while also adding new $H^2$ and $H^3$ interaction vertices with the charged leptons). Regarding the fermion masses, they are no longer diagonal due to the contribution from $(\wc_{e\phi})_{ij}$,
\begin{equation}
\Tilde{m}_{ij}^l=vy_i^l\delta_{ij}-v^3(\wc_{e\phi})_{ij}.
\end{equation}
The contribution to the Yukawa couplings is different from that of the charged lepton masses, modifying the SM prediction of simultaneous diagonalization of the fermion masses and the interactions with the Higgs boson,
\begin{equation}
v\Tilde{Y}_{ij}^l=vy^l_i\delta_{ij}  - 3v^3(\wc_{e\phi})_{ij} \neq \Tilde{m}_{ij}^l,
\end{equation}
where the general Higgs coupling to the charged leptons has been defined in Eq.~\eqref{eq:LH} which, for simplicity, we call it hereafter $Y^l\equiv Y^{(-1)}$.

The mass matrix $\Tilde{m}_{ij}^l$ is diagonalized, up to order $\mathcal{O}(1/M^2)$, through the following transformations on the LH and RH charged fermions (see \cite{delAguila:2000aa} for the corresponding
expressions in the quark sector),
\begin{align}
(U_L^l)_{ij}=\delta_{ij}+v^2(A_L^l)_{ij}, \quad (U_R^l)_{ij}=\delta_{ij}+v^2(A_R^l)_{ij},
\end{align}
with $A_L^l$ and $A_R^l$ antihermitian matrices given by
\begin{align}
    &(A_L^l)_{ij}=\left(1-\frac{1}{2}\delta_{ij}\right)\frac{y_i^l(\wc_{e\phi})_{ij}^\dagger+(-1)^{\delta_{ij}}(\wc_{e\phi})_{ij} y_j^l}{(y_i^l)^2-(-1)^{\delta_{ij}}(y_j^l)^2}, \\ 
    &(A_R^l)_{ij}=\left(1-\frac{1}{2}\delta_{ij}\right)\frac{y_i^l(\wc_{e\phi})_{ij}+(-1)^{\delta_{ij}}(\wc_{e\phi})_{ij}^\dagger y_j^l}{(y_i^l)^2-(-1)^{\delta_{ij}}(y_j^l)^2}.
\end{align}
The physical mass matrix, $m_{ij}^{phys}$, reads,
\begin{equation}
m_{ij}^{phys}=m_i^{phys}\delta_{ij}=\left(vy_i^l-\frac{v^3}{2}(\wc_{e\phi}+\wc_{e\phi}^\dagger)_{ii}\right)\delta_{ij}=m_i\left(1-\left(\frac{m_i'}{M}\right)^2\right)\delta_{ij}.
\end{equation}
In the physical basis, the Yukawa couplings are not diagonal, inducing flavor changing interactions with the Higgs boson,
\begin{equation}
\begin{aligned}
    vY^l_{ij}&=vy^l_i\delta_{ij}  - v^3 \Big[ 2(\wc_{e\phi})_{ij}+ \frac{1}{2} \delta_{ij} (\wc_{e\phi}+\wc_{e\phi}^\dagger)_{ii}\Big]=m_{ij}^{phys}-2v^3(\wc_{e\phi})_{ij}\\[4pt]
    &=m_{i}^{phys}\delta_{ij}-2\frac{m_i(m_i')^*m_j'}{M^2}=m_{i}^{phys}\left(\delta_{ij}-2\frac{(m_i')^*m_j'}{M^2}\right),
    \end{aligned} \label{eq:yukawa_physical_dim6}
\end{equation}
where we have neglected higher orders in $m_i/M$.

From Eq.~\eqref{eq:yukawa_physical_dim6} we see that, in the flavor-diagonal case, the precision with which the corresponding Yukawa coupling is experimentally measured sets directly the limit on the allowed mixing. Only the tau lepton Yukawa coupling has been measured with a certain degree of precision. Non-diagonal couplings can also induce lepton-flavor violating processes, which are very strongly constrained experimentally. However, the lepton mass suppression is enough to make these bounds irrelevant in our case. Let us review both cases here. 
In the diagonal case, the modified Yukawa coupling is proportional to the SM result. Thus, the so-called kappa framework, that assumes SM-like couplings with a modified strength, fully captures the corrections. The kappa parameter for a lepton Yukawa coupling can be defined as follows
\begin{equation}
\kappa_l \equiv \frac{Y^l}{y_l^{\mathrm{SM}}} = 1- \sin^2\theta_l,    
\label{eq:kyukawa}
\end{equation}
where in the second equality we have given the result in our model. Only the tau Yukawa coupling has been determined with enough precision to make this measurement relevant~\cite{ATLAS:2022vkf}
\begin{equation}
    \kappa_\tau = 0.93 \pm 0.07,\quad 68\%\mathrm{C.L.}.
\end{equation}
Using conservatively the error in the experimental measurement as the maximum deviation in our model we get the following $95\%$ C.L. limit,
\begin{equation}
    \sin^2{\theta_\tau} \lesssim 0.14, \label{eq:limit_tree_tau}
\end{equation}
which measures the percentage of the non-singlet component in the RH tau. 
This limit might seem independent of $M$ but, as we will see below, the interplay between tree-level and one-loop effects, makes it valid only for relatively low values of $M$.

Regarding Higgs-mediated tree-level lepton flavor violating observables, we can obtain the corresponding bounds by using the general results in Appendix~\ref{app:LFV}. Indeed, inserting the value of the relevant WCs, Eqs.~\eqref{eq:Cee} and~\eqref{eq:Cqq} in the general expressions of the corresponding LFV observables, Eqs.~\eqref{br:meee} and~\eqref{Gconv}, and using the experimental limits on Table~\ref{tab:LFVbr}, we obtain the following limits at 90$\%$ C.L.:
\begin{align}
{\rm BR}(\mu \textrm{Au}\to e \textrm{Au})&=1.7\cdot10^{-10}\left(\frac{m_\mu'm_e'}{M^2}\right)^2\leq7\cdot10^{-13}   , 
\label{eq:mue:tree}
\\
{\rm BR}(\mu\to 3e)&=4.5\cdot10^{-18}\left(\frac{m_\mu'm_e'}{M^2}\right)^2 \leq10^{-12}     ,    
\label{eq:mto3e:tree}
\\
{\rm BR}(\tau\to 3\mu)&=9.6\cdot10^{-12}\left(\frac{m_\tau'm_\mu'}{M^2}\right)^2 \leq 1.9 \cdot10^{-8}  ,   
\label{eq:tauto3e:tree}
\\
{\rm BR}(\tau\to e \mu\mu)&=9.6\cdot10^{-12}\left(\frac{m_\tau'm_e'}{M^2}\right)^2 \leq 2.7 \cdot10^{-8}.
\label{eq:tautoemumu:tree}
\end{align}

\begin{table}[h!]
\centering
\renewcommand{\arraystretch}{1.6}
\begin{tabular}{|l|l|}
\hline
$BR(\mu \to 3e) \leq 10^{-12}$~\cite{SINDRUM:1987nra} &
$BR(\tau \to 3e) \leq 2.7 \cdot 10^{-8}$~\cite{Hayasaka:2010np}  \\

$BR(\mu \to e\gamma) \leq 3.1 \cdot 10^{-13}$~\cite{MEGII:2023ltw} &
$BR(\tau \to e\gamma) \leq 3.33 \cdot 10^{-8}$~\cite{BaBar:2009hkt} \\

$BR(\mu \mathrm{Au} \to e \mathrm{Au}) \leq 7 \cdot 10^{-13}$~\cite{SINDRUMII:2006dvw} &
$BR(\tau \to \mu\gamma) \leq 4.2 \cdot 10^{-8}$~\cite{Belle:2021ysv} \\

$BR(\tau \to 3\mu) \leq 1.9 \cdot 10^{-8}$~\cite{Belle-II:2024sce} &
$BR(\tau \to \mu e e) \leq 1.8 \cdot 10^{-8}$~\cite{Hayasaka:2010np} \\
\hline
\multicolumn{2}{|c|}{$BR(\tau \to e\mu\mu) \leq 2.7 \cdot 10^{-8}$~\cite{Hayasaka:2010np}} \\
\hline
\end{tabular}
\caption{Experimental upper bounds, at $90\%$ C.L., on lepton flavor violating processes.}
\label{tab:LFVbr}
\end{table}

We note that, since the bounds in Eqs.~\eqref{eq:mue:tree}--\eqref{eq:tautoemumu:tree} arise from Higgs-mediated
tree-level processes, radiative decays involving a photon in the final state (such as $\ell_i\to\ell_j\gamma$) are 
included in the table but not in our limits, as they first appear at the one-loop level. 
Among the remaining purely tree-level LFV processes, we only consider the most constraining observable for each flavor transition
($\mu\!-\!e$, $\tau\!-\!\mu$, and $\tau\!-\!e$), which explains why the number of relations in
Eqs.~\eqref{eq:mue:tree}--\eqref{eq:tautoemumu:tree} is smaller than the total number of experimental limits listed in Table~\ref{tab:LFVbr}.

The numerical factor on the LHS of each inequality corresponds to the replacement of the corresponding masses in the expressions of the observables and the RHS corresponds to the experimental limit, as shown in Table~\ref{tab:LFVbr}. %As we can see 
Among the bounds on that table, only $\mu-e$ conversion imposes a non-negligible limit, 
\begin{equation}
   \frac{m_\mu'm_e'}{M^2} \leq 0.065. 
   \label{eq:limit_mue_tree}
\end{equation}
In all other cases the light-lepton mass suppression is enough to make the bounds meaningless. As we will see in Section~\ref{sec:LFV_oneloop}, even the constraint in Eq.~\eqref{eq:limit_mue_tree} becomes surpassed by the LFV constraints arising at one-loop order.

\subsection{Tree-level contributions to the dimension 8 SMEFT \label{sec:tree-level-dim8}}
To assess whether tree-level dimension-8 effects are relevant, let us consider only their possible contributions to EWPO and Higgs physics that were absent at tree-level in the dimension 6 SMEFT. Even considering that these corrections appear at one-loop at dimension-6, dimension-8 tree-level contributions could, in principle, be relevant for smaller values of $M$. 

It is clear that any dimension 8 operator that we generate with a coefficient suppressed by the charged lepton Yukawa couplings will be too small to be observable. Thus, we take the vanishing Yukawa limit to simplify the discussion. In the following we will show that none of the operators generated at this order (in the vanishing Yukawa limit) give rise to 3-point  interactions and, therefore, they do not affect EWPO. Using \texttt{Matchete}~\cite{Fuentes-Martin:2022jrf} for the integration we obtain the following dimension 8 contribution
\begin{align}
    \mathcal{L}_{\textrm{SMEFT}}&\supset\left(-\frac{\mu^2}{M^4}+\frac{\lambda}{M^4} \phi^\dagger \phi\right)(\phi^\dagger \ii \overset\leftrightarrow{D_\mu} \phi\, \overline{e}\gamma^\mu e) - \frac{\ii}{M^4}\left(D_\mu\phi^\dagger D^\mu D^\nu\phi - D_\mu D^\nu\phi^\dagger D^\mu \phi\right) \overline{e}\gamma_\nu e \nonumber\\
    &+ \frac{2\ii}{M^4} D_\mu\phi^\dagger D_\nu \phi\left(\overline{e}\gamma^\mu D^\nu e - D^\mu\overline{e}\gamma^\nu e \right) + \frac{g_1^2}{4M^4} D^\mu (\phi^\dagger \phi) B_{\mu\nu} \overline{e}\gamma^\nu e  + \dots,
    \label{eq:L8}
\end{align}
where the dots correspond to further operators in the class $\psi^2XH^2D$ with the same structure of the last operator but more gamma matrices. 
It might seem that the first line in Eq.~\eqref{eq:L8} would give rise to  
anomalous couplings of the gauge bosons to the fermions. 
However, removing the last term in the first line, through an appropriate field redefinition, cancels the contribution to $\mathcal{O}_{\phi e}$ and $\mathcal{O}_{\phi e }(\phi^\dagger\phi)$ and further cancels the first term in the second line. The leading tree-level dimension-8 contribution will therefore arise from a 4-point interaction, given by operators in the classes $\psi^2 X H^2 D$.  

An alternative way of obtaining the same result is to consider directly the theory in its broken phase (see Section~\ref{sec:model}) and integrate out the heavy degrees of freedom in the physical basis. Integrating out the VLLs in the Lagrangian of Eqs.~\eqref{eq:LZ}-\eqref{eq:LH} results in
\begin{align}
\label{eq:left}
    \mathcal{L}_{\textrm{LEFT}} &\supset \left(\frac{m^\prime}{M^2}\right)^2 \overline{e}_R \slashed{Z} (\ii\slashed \partial) \slashed{Z} e_R + \left[\sqrt{2} \left(\frac{m^\prime}{M^2}\right)^2\overline{e}_R \slashed{W}^- (\ii\slashed \partial) \slashed{W}^+ e_R + \mathrm{h.c.}\right] \nonumber\\
    &+\left(\frac{m^\prime}{v M}\right)^2 \overline{e}_R H \left[\left(1-\frac{\partial^2}{M^2}\right)\right](\ii\slashed{\partial})H e_R\,,
\end{align}
where we suppressed flavor indices for simplicity. Upon rewriting the dimension-6 part of the last term in an explicitly hermitian form, 
\begin{align}
    \overline{e}_R H(\ii\slashed{\partial})H e_R &= \frac12 \left[\overline{e}_R H (\ii \slashed{\partial}) (H e_R) - (\ii \slashed{\partial}) (\overline{e}_R H) He_R\right] \nonumber \\
    &= \frac12 \left[ 2 H^2 \overline{e}_R\ii\slashed{\partial}e_R + (H\partial_\mu H-H\partial_\mu H) \ii \overline{e}_R\gamma^\mu e_R \right]\,,
\end{align}
it becomes clear that it results only in a lepton mass suppressed contribution, as expected from the SMEFT results.

From Eq. \eqref{eq:left}, we recover the same conclusion from the unbroken SMEFT: the effective theory points to the $e^+ e^-\xrightarrow{}WW/ZZ$ processes as the more relevant dimension-8 tree-level contributions, as no 3-point vertices are generated at this order. Further reductions of Eq. \eqref{eq:left} (for instance through Fierz reordering) to arrive at a minimal basis are not necessary at this stage, as henceforth we will consider the full model in the computation of the $e^+ e^-\xrightarrow{}WW/ZZ$ predictions. These interactions were probed by LEP (see for instance~\cite{Badaud:2000qw}) but, as we will see, they are not competitive with other current constraints. For the sake of concreteness, we have considered the results in~\cite{L3:2004lwm} for $W^+ W^-$ production in $e^+ e^-$ collisions, as they are the most precisely measured. Instead of simulating in detail the corresponding analyses we have taken the very conservative approach to put a limit on the model when the production cross-section in our model differs by more than $0.1$ pb with the one in the SM. This difference is much smaller than the precision of experimental data, in some cases by more than one order of magnitude. We will show the corresponding limit in our final plot in Figure~\ref{fig:excl_global_e}, where we will see that it is completely irrelevant when direct searches and theoretical limits are considered

\subsection{One loop contributions to the dimension 6 SMEFT \label{sec:one-loop-dim6}}

Let us finally turn our attention to the effect of dimension-6 operators with WCs generated at one-loop order. We have integrated out the two heavy multiplets up to one loop order using \texttt{Matchete}~\cite{Fuentes-Martin:2022jrf} and \texttt{MatchmakerEFT}~\cite{Carmona:2021xtq} and checked that both calculations agree. The full result is too long to be given in writing. We collect some of the most relevant contributions in Appendix~\ref{appendix:oneloop:matching} and give the full result as an ancillary file in the arXiv submission of this work. Given that, in the case that the heavy leptons mix with more than one light fermion, we generate lepton-flavor-violating (LFV) operators at one loop order that are not suppressed by light Yukawa couplings, we first consider the constraints that these LFV observables impose and then we consider the lepton flavor preserving constraints.

\subsubsection{Constraints from Lepton Flavor Violating Observables \label{sec:LFV_oneloop}}
\begin{figure}[h!]
\centering
\begin{subfigure}{.6\textwidth}
  \centering
  \includegraphics[width=.8\linewidth]{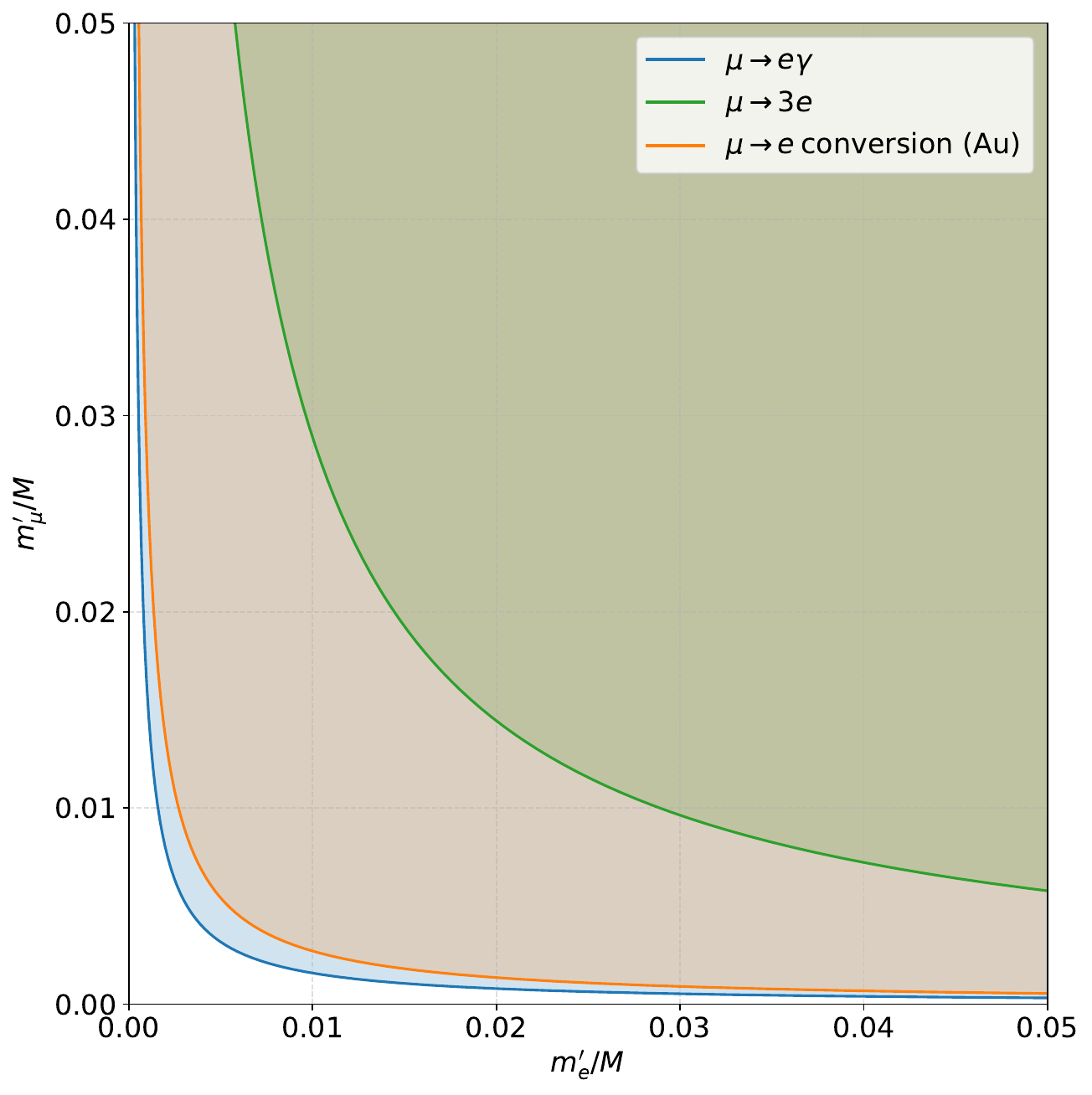}
\end{subfigure}
\\
\begin{subfigure}{\textwidth}
  \centering
  \includegraphics[width=.45\linewidth]{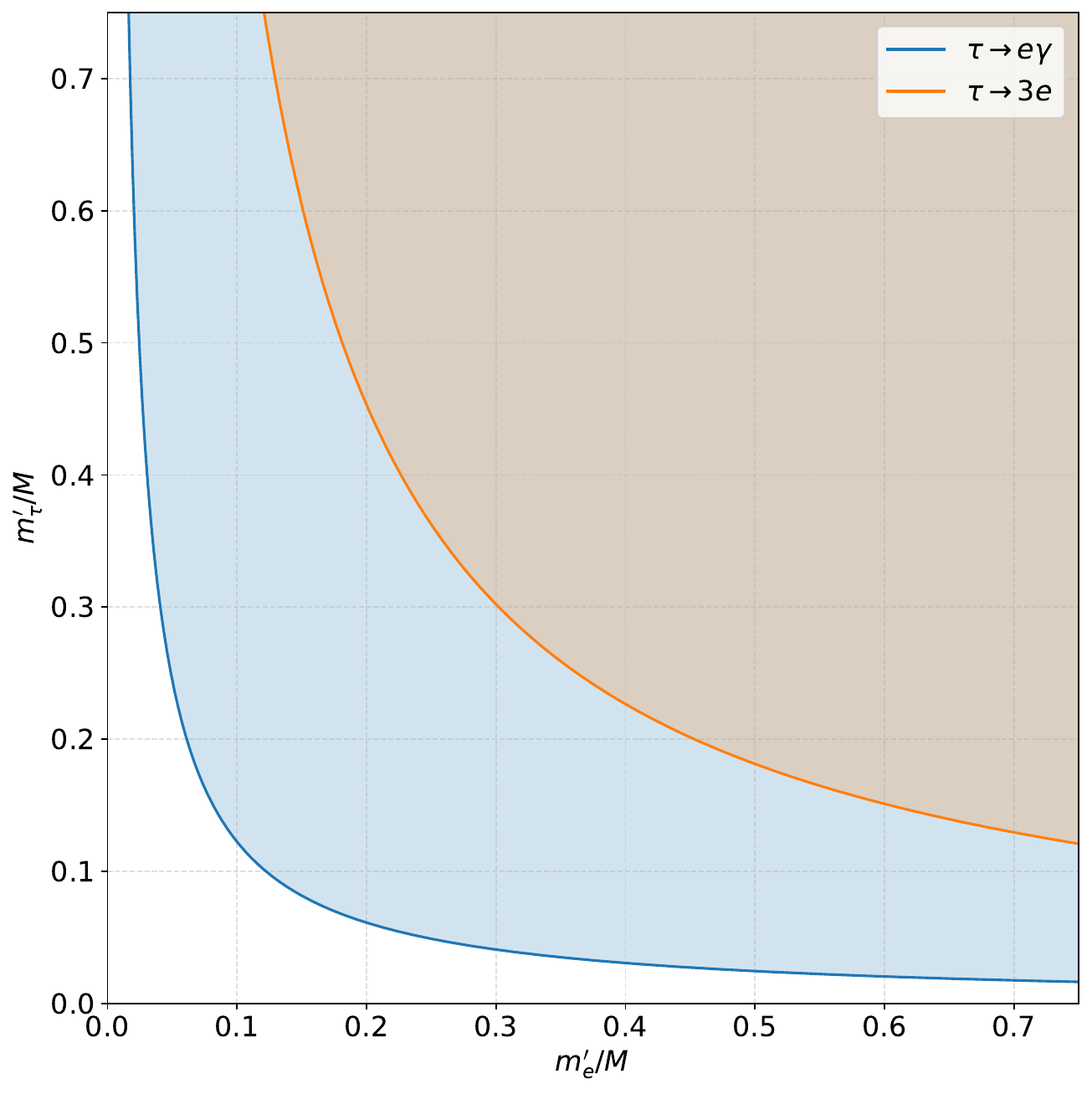}
  \includegraphics[width=.45\linewidth]{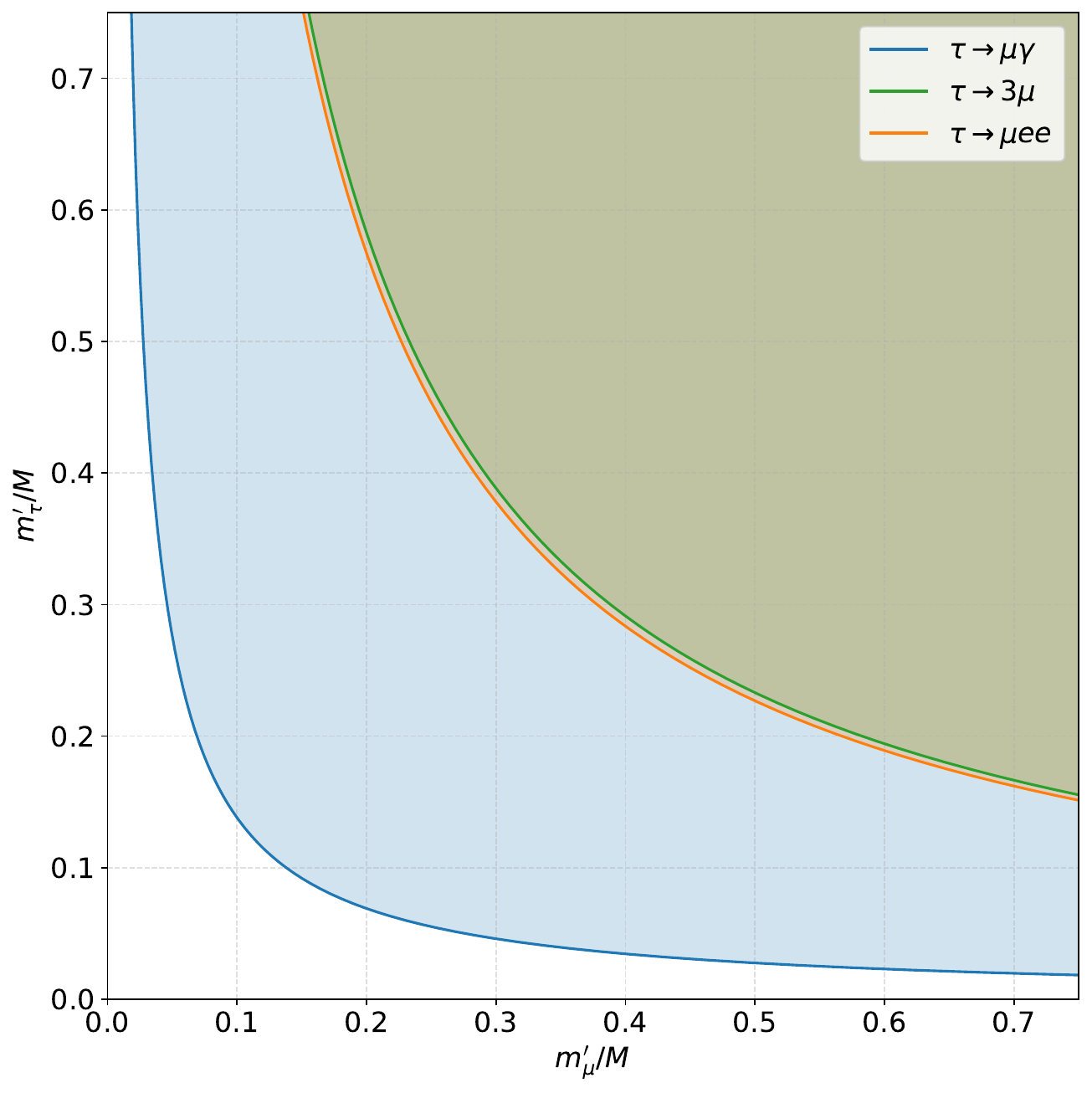}
\end{subfigure}
\caption{90\% C.L. LFV limits on $\frac{m^\prime_i m^\prime_j}{M^2}$ with $i\neq j$ for $\mu-e$ transitions (top panel), $\tau-e$ transitions (bottom left panel) and $\tau-\mu$ transitions (bottom right panel). }
\label{fig:LFV_limits}
\end{figure}

When our VLLs mix with several different light generations, \textit{i.e.} there are at least two non-vanishing $\lambda^\prime_{i,j}$ with $i\neq j$, LFV processes will be induced at one loop order. Interestingly enough, the symmetry that protects the tree-level contributions ensures a partial cancellation of these effects, so that $\Delta L=2$ operators, proportional to $(\lambda^{\prime \,\ast}_i)^2 (\lambda^\prime_j)^2$, are zero in our model. 
This results from a cancellation between box diagram contributions involving the two VLL doublets. Although each doublet individually generates such contributions, they have the same size and opposite signs, so that they cancel exactly. Nevertheless, there are $\Delta L=1$ processes proportional to $\lambda^{\prime \, \ast}_i \lambda^\prime_j$ that, as we will see, imply very stringent constraints on this product. 
All the relevant experimental observables are provided in Table~\ref{tab:LFVbr}. These observables can be computed in terms of the parameters of our model $\lambda^\prime_i$ and $M$. Due to the particular symmetry protection of the model, they solely depend on the product $\lambda^{\prime\, \ast}_i \lambda^\prime_j/M^2$ for $l_i \to l_j$ processes (or on its complex conjugate, depending on the electric charge of the leptons used). The procedure, described in detail in Appendix~\ref{app:LFV}, consists of matching our SMEFT results to the LEFT and then computing the relevant observables in terms of the WCs in the LEFT. We assume that the couplings $\lambda^\prime_i$ are real, but this assumption can be trivially lifted knowing the corresponding dependence on the parameters. The resulting limits, graphically shown in Figure~\ref{fig:LFV_limits}, are given, in decreasing order of importance, by
\begin{align}
\frac{m^\prime_\mu m^\prime_e}{M^2} \leq  
&\left\{
\begin{array}{l}
1.58 \cdot 10^{-5}, \quad [\mu \to e \gamma], \\
2.70 \cdot 10^{-5}, \quad [\mu \to e \text{ conversion in Au}],\\
2.89 \cdot 10^{-4}, \quad [\mu \to 3 e],
\end{array}
\right. \\[6pt]
\frac{m^\prime_\tau m^\prime_e}{M^2} \leq  
&\left\{
\begin{array}{l}
0.012, \quad [\tau \to e \gamma], \\
0.091, \quad [\tau \to 3 e],
\end{array}
\right.
\end{align}
\begin{align}
\frac{m^\prime_\tau m^\prime_\mu}{M^2} \leq  
\left\{
\begin{array}{l}
0.014, \quad [\tau \to \mu \gamma], \\
0.113, \quad [\tau \to \mu e e], \\
0.116, \quad [\tau \to 3 \mu].
\end{array}
\right.
\end{align}
As announced in Section~\ref{sec:tree-level-dim6} these limits clearly supersede the ones obtained at tree level.

\subsubsection{Flavor preserving global fit}
\label{sec:flavourpreservingfit}

Given the tight constraints on the simultaneous mixing with two different light leptons discussed in the previous section, we turn our attention to the lepton flavor preserving case and perform a global fit to electroweak precision measurements, including the anomalous magnetic moment of the muon, and Higgs-boson observables at the LHC. 
We have used the python package \texttt{smelli}~\cite{Aebischer:2018iyb,Straub:2018kue,Aebischer:2018bkb} to compute the global likelihood in terms of the parameters of our model, $\lambda^\prime$ and $M$, using the matching results of our model onto the SMEFT -- see Appendix~\ref{appendix:oneloop:matching}. 
We note that most of the effects that enter in the observables considered in the fit are induced by one-loop generated WCs. 
In this case, to maintain a consistent one-loop power counting, the effects from renormalization group evolution (RGE) should be neglected, as they induce corrections of the next order in perturbation theory. The exception to this argument concerns the WCs of the operators that modify the Yukawa couplings of the leptons. These are generated at the tree level, and are particularly relevant for the case of mixing with the tau. However, even in this case, we checked that the effects of RGE do not significantly change the results, and are neglected in what follows.

\begin{figure}[h!]
\centering
  \includegraphics[width=.49\linewidth]{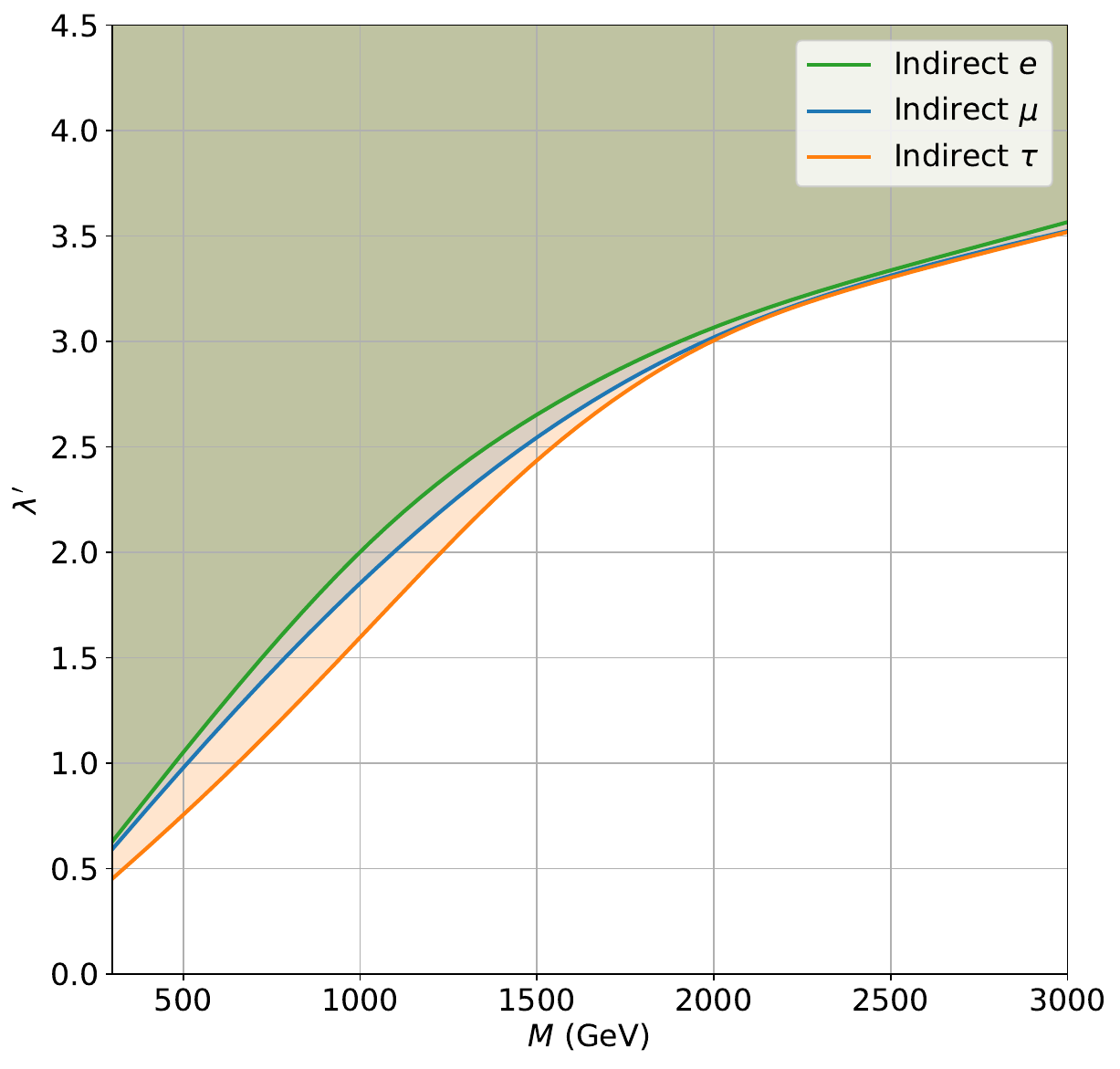}
 \includegraphics[width=.49\linewidth]{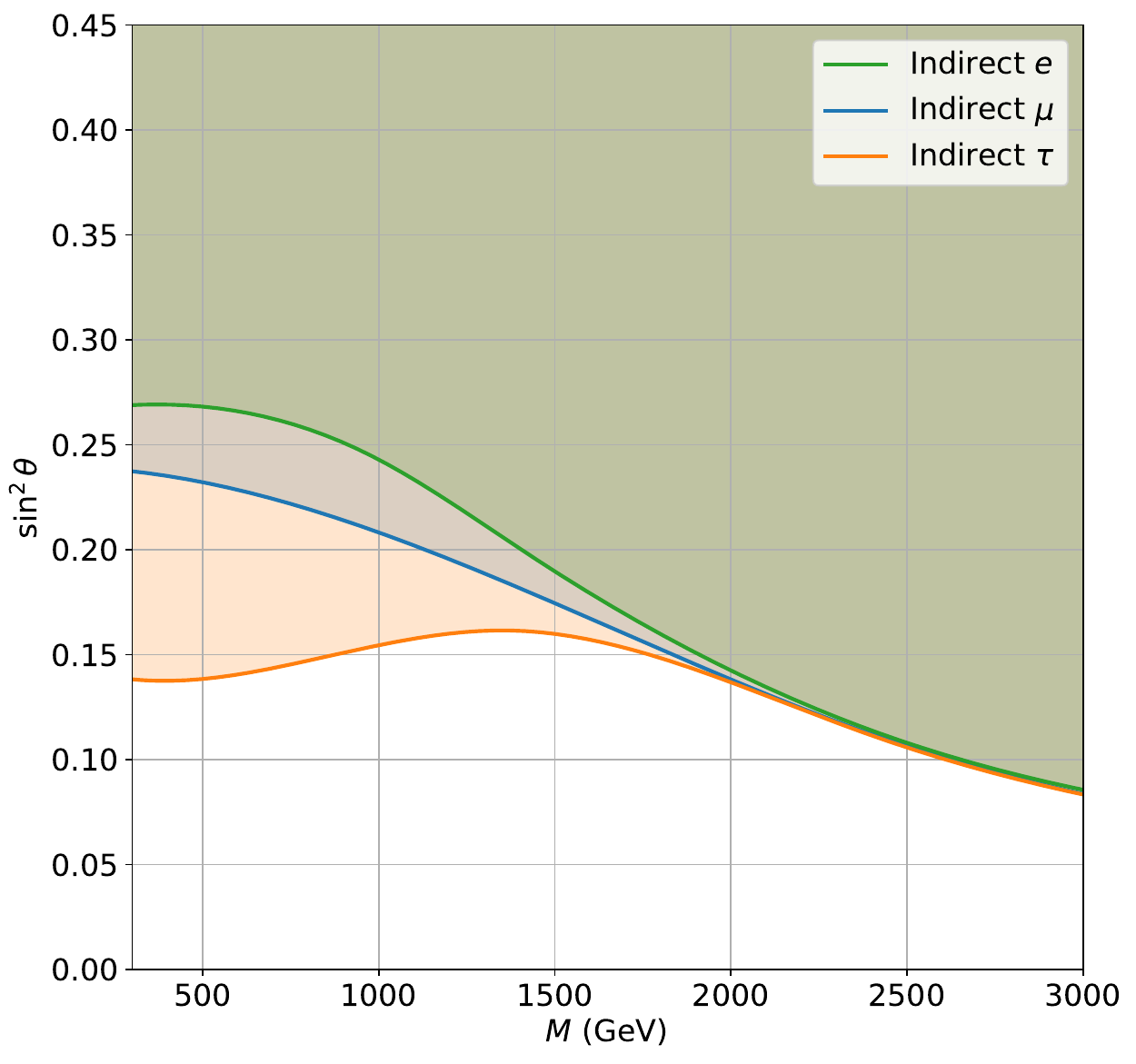}
  \label{fig:combination_e_mixing}
\caption{Excluded regions (colored areas) at 95\% C.L. in the $\lambda^\prime-M$ plane (left panel) and in the $\sin^2\theta-M$ plane, for single generation mixing with $e$ (green), $\mu$ (blue) and $\tau$ (orange), respectively.}
\label{fig:global_fit_all}
\end{figure}

Once we have the relevant $\chi^2$ as a function of $\lambda^\prime$ and $M$~\footnote{Note that the one-loop WCs have contributions with dependence on different combinations of these parameters, including $\lambda^{\prime\,2}/M^2$, $\lambda^{\prime\,4}/M^2$ and $\lambda^{\prime\,6}/M^2$ (see Appendix~\ref{appendix:oneloop:matching}) so the likelihood depends on both parameters independently.} we can obtain the $95\%$ C.L. limits on the $M-\lambda^\prime$ plane by considering the excluded regions that satisfy $\Delta\chi^2=\chi^2-\chi^2_{\textrm{min}}\geq5.99$.
The most constraining observables are EWPO and Higgs physics, except for the case of mixing with the muon, for which $g-2$ plays an important role. 
In fact, Higgs physics, and in particular double Higgs production, $H\to \gamma \gamma$ and $H\to \tau \tau$ (for mixing with the tau), induces the strongest constraints, with EWPO being relevant mostly for smaller masses. We illustrate our findings for the three cases of mixing with $e$, $\mu$ and $\tau$, respectively, in Figure~\ref{fig:global_fit_all}. The exclusion regions are given in the $\lambda^\prime-M$ plane in the left panel of the figure and in the $\sin^2\theta-M$ plane in the right one. 

Let us focus on the right panel of Figure~\ref{fig:global_fit_all}, as the dependence of the different observables on the mixing squared is more easily identifiable. The three curves behave differently until they merge into each other for larger masses, at $M\sim1.5$ TeV. 
Below this mass, $M\lesssim 1.5$ TeV, the limits for the case of mixing with the electron and the muon look similar, with two lines that are relatively close to each other. However, the origin of the constraints is slightly different in each case. They share a common leading constraint, from the measurement of the $H \to \gamma \gamma$ decay. 
This decay receives a large contribution from $\wc_{\phi B}$ that makes it the dominant channel for small masses. EWPO impose a milder but non-negligible constraint, which is more relevant for the electron case. In the muon case, the measurement of the anomalous magnetic moment of the muon, $g-2$, places very significant constraints, rivaling the one arising from $H\to \gamma \gamma$ at small values of $M$. All these observables depend only on WCs that scale like $\lambda^{\prime\,2}/M^2$ resulting in an $M$-independent constraint on the mixing (see Appendix~\ref{appendix:oneloop:matching} for the WCs and Appendix~\ref{app:gminus2} for $g-2$ calculation). 
For a fixed value of the mixing, as $M$ (or $\lambda'$) grows the contributions to $C_{\phi}$ and, to a less extent, $C_{\phi \Box}$, become increasingly important, since they scale like $\lambda^{\prime\,6}/M^2$ and $\lambda^{\prime\,4}/M^2$, respectively. 
The operator $\wc_{\phi}$ only corrects the Higgs self-coupling $\kappa_\lambda$ (see Appendix D for the calculation of $\delta\kappa_\lambda$ in our model). 
Despite the comparatively lower precision of the bounds from Higgs pair production, the scaling of this WC makes this observable the leading constraint, and causes these curves 
to acquire a downward slope for large values of $M$.  

Limits for the mixing with the tau look quite different, since they originate from the constraint on the tau Yukawa coupling to the Higgs boson, which has been measured with a precision at the level of several percent, making the tree level contribution to this coupling relevant for low values of $M$, see Eq.\eqref{eq:limit_tree_tau}. As $M$ increases (and equivalently $\lambda'$ increases for the same value of the mixing), there is a partial cancellation with the one-loop result, explaining the slight upwards slope of the bound; we will further discuss this one-loop cancellation in section~\ref{sec:theory_bounds}. 
For larger masses, $M\gtrsim1.5$ TeV, double Higgs production becomes again the leading constraint, turning it into the dominant observable for the three scenarios, merging the three lines together.

\begin{figure}[t!]
\centering
  \includegraphics[width=.32\linewidth]{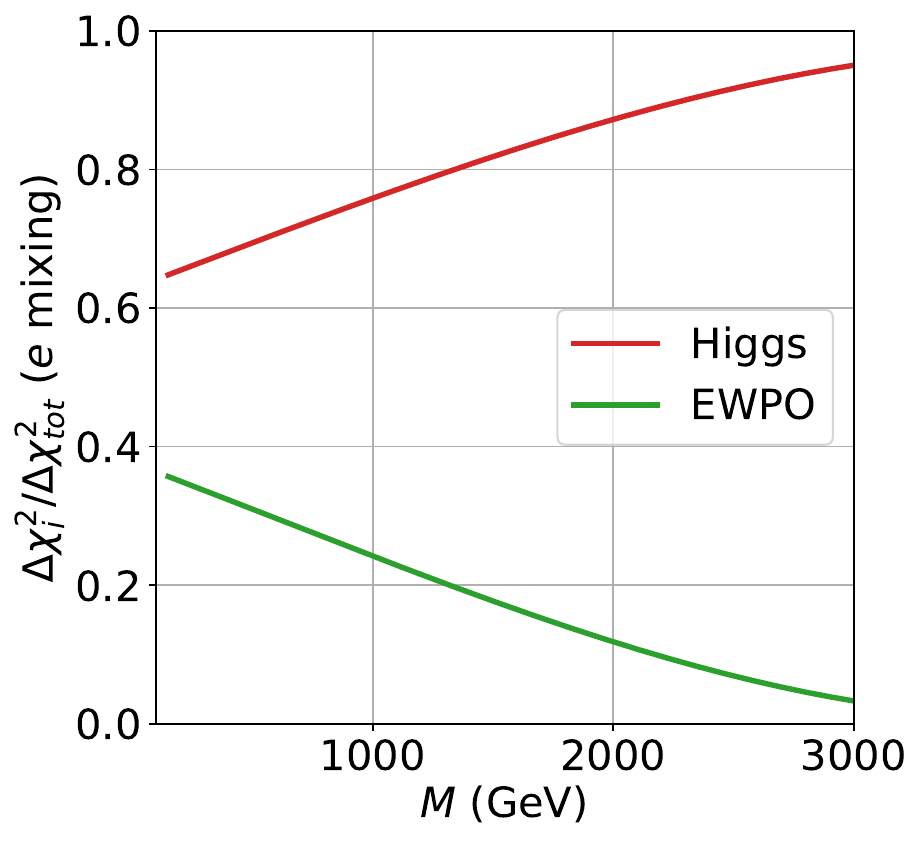}
  \includegraphics[width=.32\linewidth]{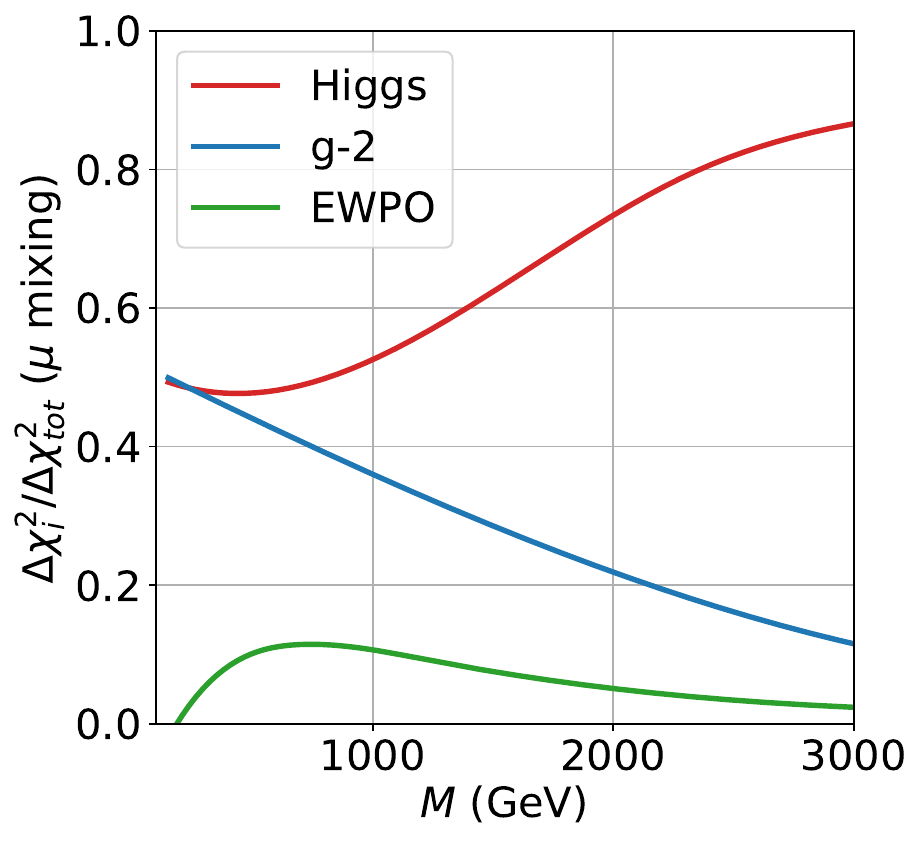}
 \includegraphics[width=.32\linewidth]{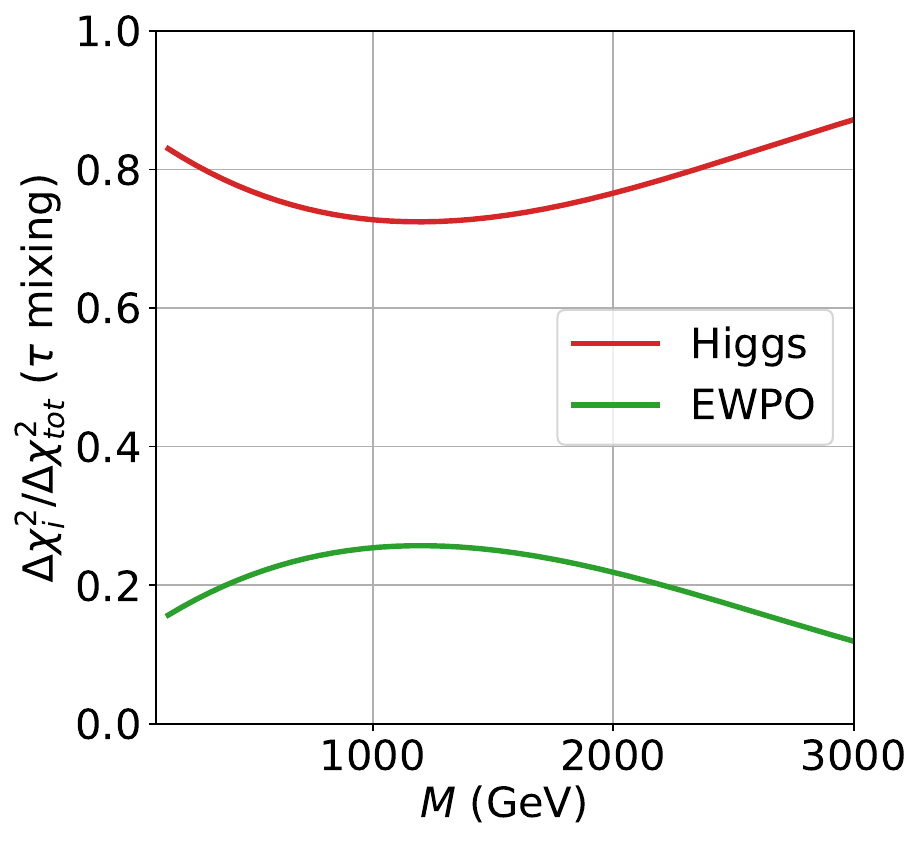}
\caption{Fraction of the total $\Delta\chi^2$ corresponding to EWPO (green), Higgs observables (red) and $g-2$ (blue, only relevant in the muon case) in the exclusion limit for the case of mixing with electron (left), muon (center) and tau (right).}
\label{fig:chi_comparison_e_mu_tau}
\end{figure}

The features mentioned above can be easily understood from the fraction of the $\Delta \chi^2=5.99$ that arises from the different sets of (uncorrelated) observables. We show these fractions in Figure~\ref{fig:chi_comparison_e_mu_tau} for the case of mixing with electron (left), muon (center) and tau (right), respectively. We see how, for the electron, Higgs physics is dominant, with EWPO playing a smaller but still significant role for small values of $M$. In the muon case we see that the constraint from $g-2$ is relevant, being comparable to Higgs constraints for low values of $M$. The $\lambda^{\prime\,4}/M^2$ and $\lambda^{\prime\,6}/M^2$ behavior of Higgs physics makes it more relevant for larger values of $M$ against the $M$-independent contribution from other observables. Finally, in the case of mixing with the tau Higgs physics leads the bound due to the tree-level correction to the tau Yukawa coupling and to double Higgs production for large values of $\lambda'$. 

\section{Theoretical constraints~\label{sec:theory_bounds}}

The indirect constraints on the mixing discussed in the previous sections are so mild that, when combined with the relatively stringent limit on the mass from direct searches, they cast doubts on the theoretical validity of the corresponding region of parameter space. In this section we discuss different theoretical constraints on the parameters of the model. We consider three different types of theoretical limits: 
the stability of the Higgs potential, existence of Landau poles and strong coupling.

\subsection{Stability of the scalar potential}

The large values of the coupling $\lambda^\prime_i$ that experimental probes currently allow can affect the stability of the EW vacuum,  
inducing a theoretical limit on this parameter. Let us consider the Higgs potential up to mass-dimension 8,
\begin{equation}
    V(\phi)= c_2 \,(\phi^\dagger \phi) + c_4 (\phi^\dagger \phi)^2
     + c_6 (\phi^\dagger \phi)^3
      + c_8 (\phi^\dagger \phi)^4 + \ldots~.
\end{equation}
The leading contributions to the different coefficients are, assuming mixing with a single generation induced by a real coupling $\lambda^\prime$,
\begin{align}
c_2 &\sim - |\mu_H|^2 + \frac{\lambda^{\prime\,2} M^2}{4 \pi^2} \to - |\mu_H|^2, \label{eq:c2}
\\
c_4 &\sim \lambda - \frac{5}{24 \pi^2} \frac{\lambda^{\prime\,4}|\mu_H^2|}{M^2},
\\
c_6 &\sim - \frac{\lambda^{\prime\,6}}{6\pi^2M^2},
\\
c_8 &\sim \frac{\lambda^{\prime\, 8}}{12\pi^2 M^4}.
\end{align}
In the expressions above we have only considered the leading contribution in the large $\lambda^\prime$ limit. In Eq.~\eqref{eq:c2} we find the usual effect of the hierarchy problem, which is irrelevant in our case because all effects that come with $\mu_H$ appear at one-loop. Note that the contribution to $c_6$ is negative, thus destabilizing the potential, which is only stabilized at large values of $\phi$ by the (positive) $c_8$ term.

This potential depends on four input parameters: $\mu_H$, $\lambda$, $\lambda^\prime$ and $M$. As we are interested in a scan over the latter two, for each value of the pair $\lambda^\prime$, $M$, we fix $\mu_H$ and $\lambda$ by requiring that the potential has a minimum at $|\phi|^2=v^2$, with $v=174$ GeV (note that in our model, at the order we are working, there are no further contributions to the Fermi constant), and the mass of the physical Higgs boson is $m_H=125$ GeV. In order to do that, we expand~\cite{Hays:2018zze}
\begin{equation}
\phi= \begin{pmatrix}
    0 \\ v+(1+c_{H,\mathrm{kin}})\frac{H}{\sqrt{2}}
\end{pmatrix},
\end{equation}
where the contribution from canonically normalizing the Higgs is, in our model (dimension-8 contributions cancel in our case),
\begin{equation}
    c_{H,\mathrm{kin}}= - \frac{ \lambda^{\prime\,4} v^2}{24 \pi^2 M^2}.
\end{equation}

\begin{figure}[h!]
    \centering
    \includegraphics[width=0.5\linewidth]{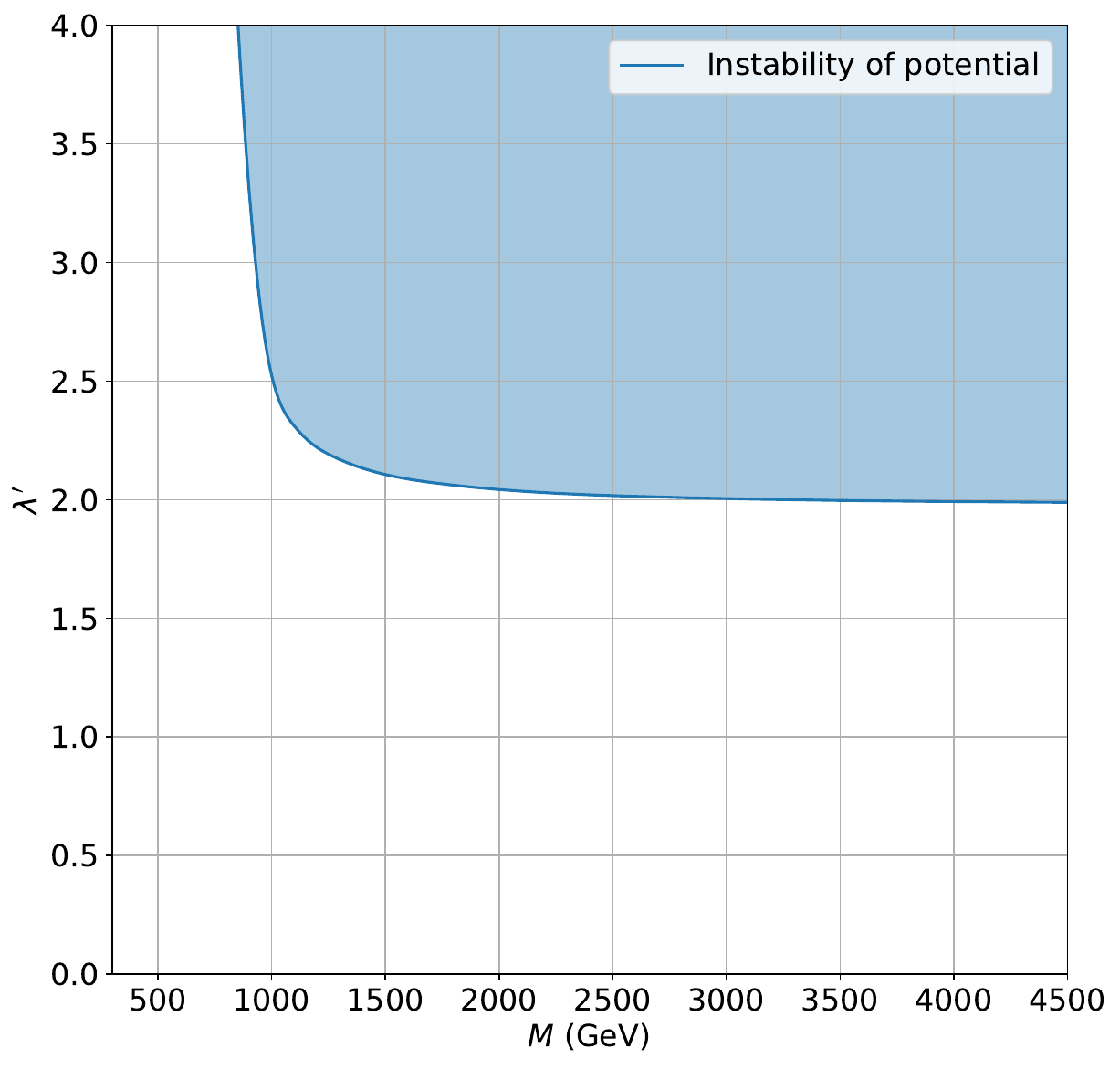}
    \caption{Maximum value of $\lambda^\prime$ as a function of $M$ from stability of the potential, see text for details.  }
    \label{fig:lampvsM_stability}
\end{figure}

For the parameters of interest in the relevant ranges, $\lambda^\prime \sim \mathcal{O}(2-4)$ and $1.5 \mbox{ TeV} \lesssim M \lesssim 4 \mbox{ TeV}$, the behavior of the potential is as follows. For fixed $M$ in the relevant region and small values of $\lambda^\prime$, the potential has a single minimum, corresponding to the observed EW one at $v=174$ GeV. As we increase $\lambda^\prime$ a second minimum develops at around $\lambda^\prime \approx 2$, with a very mild dependence on the value of $M$. 
Above this value, the second minimum becomes deeper than the observed one, thus rendering our minimum metastable. We consider this as the limiting value of $\lambda^\prime$ from the stability of the potential. We show the result in Figure~\ref{fig:lampvsM_stability}, where we see that $\lambda^{\prime} \lesssim 2$, quite independently of $M$, for the relevant range of this parameter.

As a cross-check we use the general analysis of the stability of the potential in Refs.~\cite{Falkowski:2019tft,Durieux:2022hbu}. In particular, using Eq.~(3.21) from the arXiv version of~\cite{Durieux:2022hbu}, we obtain $\lambda^{\prime} \lesssim 1.97$, a value consistent with our analysis. 

One could add bosonic degrees of freedom to the model to alleviate this theoretical constraint. The issue of potential stability for new fermions with large couplings has been addressed before in the literature, see for instance~\cite{Davoudiasl:2012tu,Blum:2015rpa,Gopalakrishna:2018uxn,DAgnolo:2023rnh,Adhikary:2024esf}.

\subsection{Landau Pole}

\begin{figure}[h!]
    \centering
    \includegraphics[width=0.5\linewidth]{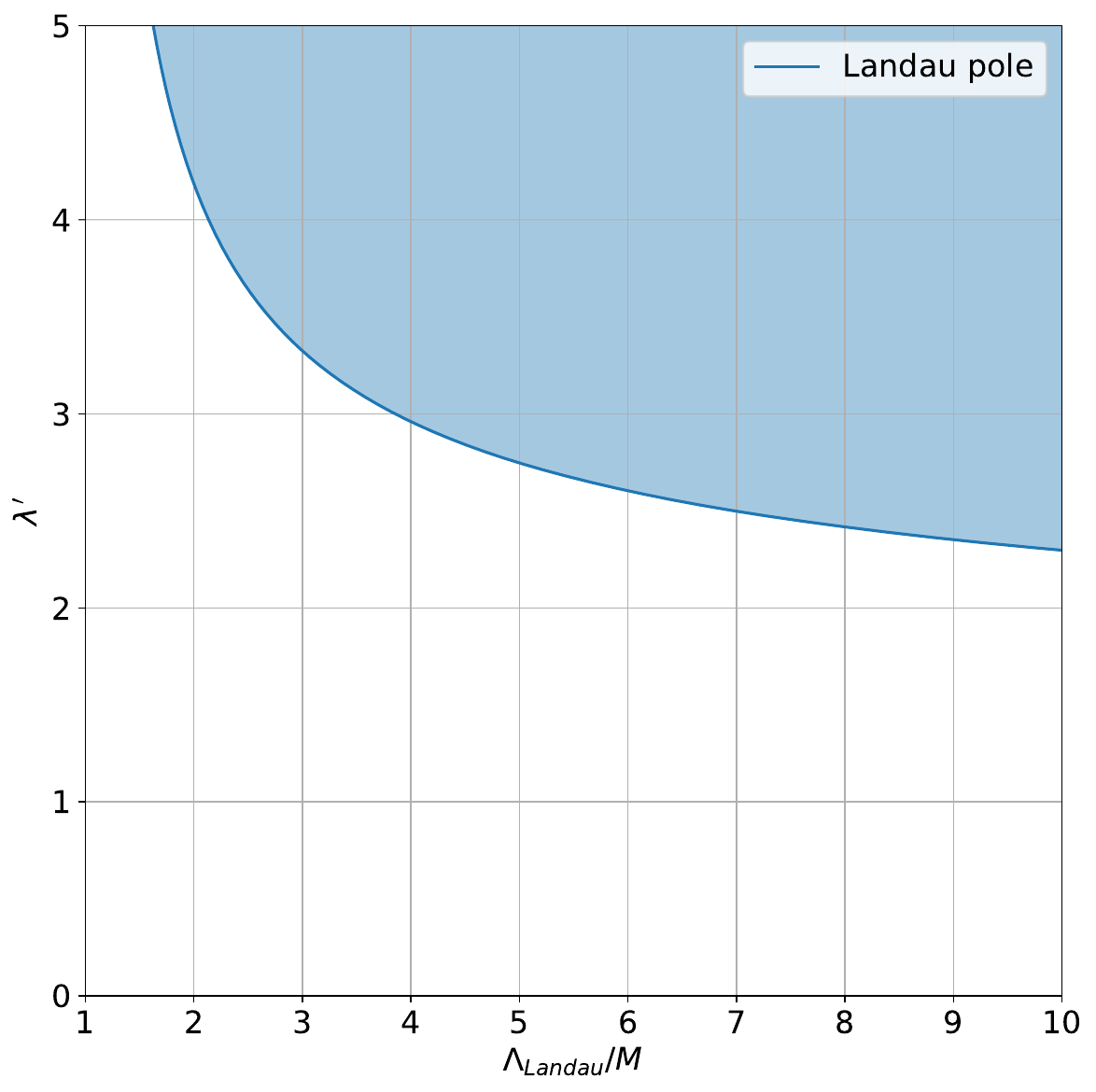}
    \caption{Excluded region in the $\Lambda_{\mathrm{Landau}}/M-\lambda^\prime$ plane from the appearance of the Landau pole at a scale $\Lambda_{\mathrm{Landau}}$ in $\lambda^\prime_{1,3}$, see text for details.  }
    \label{fig:landaupole}
\end{figure}

Another important theoretical constraint on the value of $\lambda^\prime$ arises from the presence of a Landau pole. We estimate this constraint by solving the one-loop RGEs of $\lambda_{1,3}^\prime$, Eqs.~\eqref{eq:lam1pdot} and~\eqref{eq:lam3pdot}, neglecting for simplicity the terms proportional to the gauge couplings as they are irrelevant for large values of $\lambda^\prime$, with the boundary condition $\lambda^{\prime}_{1}(M)=\lambda^{\prime}_{2}(M)=\lambda^\prime$ and fix the position of the Landau pole as the value of the renormalization scale at which $\lambda^{\prime}_{1,3}$ blow up (neglecting $g_1$ they remain equal at all scales). This gives us the value of $\lambda^\prime$ as a function of the ratio $\Lambda_{\mathrm{Landau}}/M$, shown in Figure~\ref{fig:landaupole}. As an example, for $\Lambda_{\mathrm{Landau}}/M=2, 4, 10$ we obtain, respectively $\lambda^\prime \lesssim 4.2, 3, 2.3$.

\subsection{Strong coupling}

We have estimated the strong coupling limit $\lambda^\prime \lesssim 4$ in two different ways. The first one is by computing the value of $\lambda^\prime$ for which perturbative unitarity is broken. Following~\cite{Allwicher:2021rtd}, and assuming conservatively the mixing with only one VLL, we have checked that the most stringent bound arises from diagonalizing the partial wave scattering amplitude for total angular momenta $J=1$, resulting in $\lambda^\prime \lesssim 4$, much stronger than considering only the perturbativity bound from the anomalous dimension matrix, Eq. \eqref{eq:uvrges}. 

As an alternative way of estimating the strong coupling limit, let us focus on the perturbativity of our matching results. Focusing on the matching of the $C_{e\phi}$ coefficient, which is the only one where tree- and loop-level effects compete, we have, for mixing with a single generation,
\begin{equation}
(\wc_{e\phi})_{ii} = y_i \frac{\lambda^{\prime\,2}_i}{M^2} \left[ 1 - \frac{79}{96\pi^2} \lambda^{\prime\,2}_i \right] +\ldots ~,\label{eq:tau_yukawa_cancellation}
\end{equation}
where the $\ldots$ represent other gauge coupling suppressed contributions. Immediately we observe that there is a partial cancellation between the tree- and loop-level results, explaining the behavior observed for the limit on the mixing with the tau in section~\ref{sec:flavourpreservingfit}. Furthermore, we can see that because the contributions scale differently with $\lambda'$, the one-loop result matches the tree-level one for $\lambda'=3.46$. Clearly this points to the breakdown of our perturbative expansion and we cannot be confident that higher-orders will not be important. Regardless, since the theoretical bound from vacuum stability places a more stringent limit on $\lambda'$, this will not be an issue.

\section{Global current constraints \label{sec:combination}}

\begin{figure}[h!]
    \centering
    \includegraphics[width=0.6\linewidth]{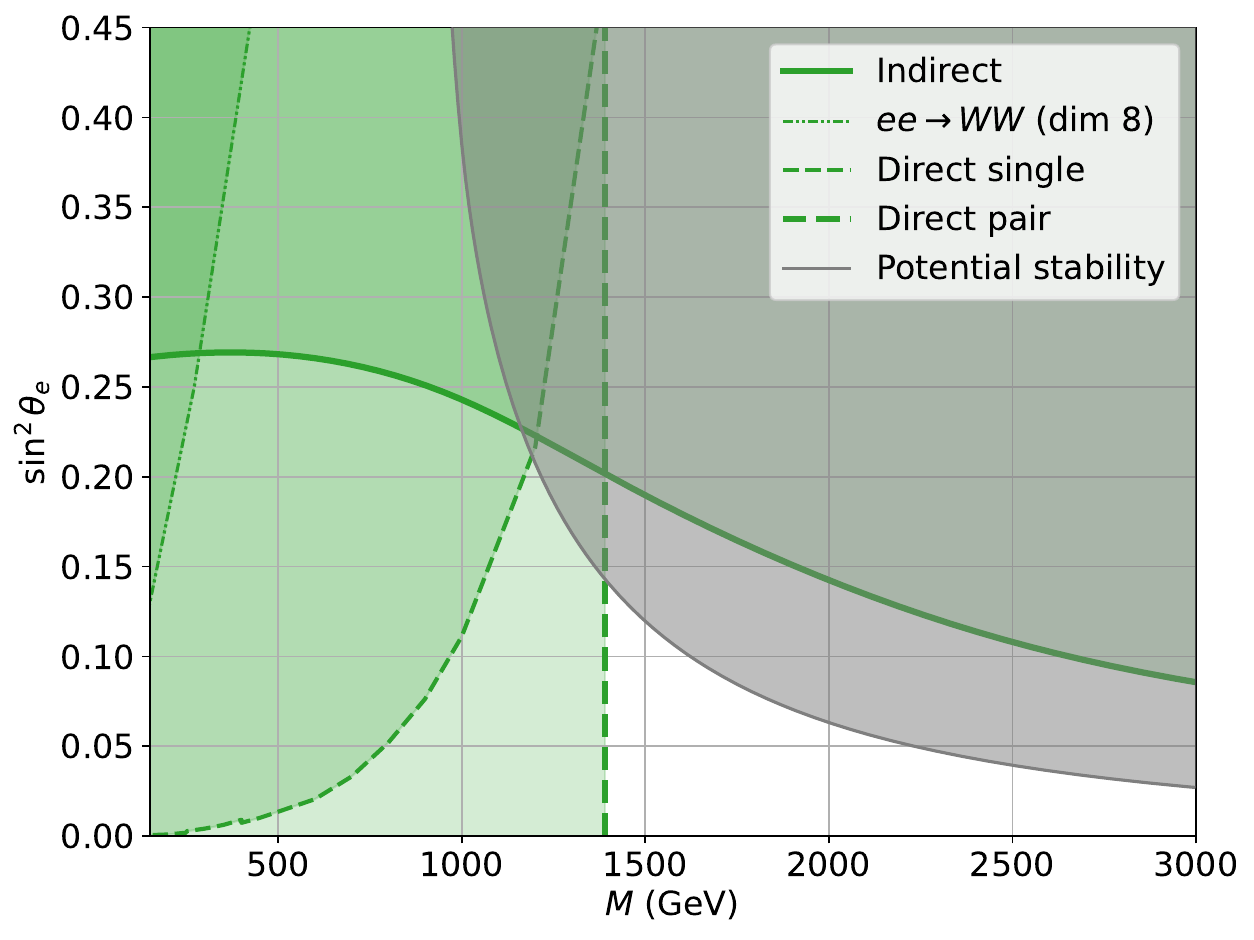}
    \caption{Combination of all exclusion regions at $95\%$ C.L. for the case of mixing with the electron. }
    \label{fig:excl_global_e}
\end{figure}

\begin{figure}[t!]
    \centering
    \includegraphics[width=0.48\linewidth]{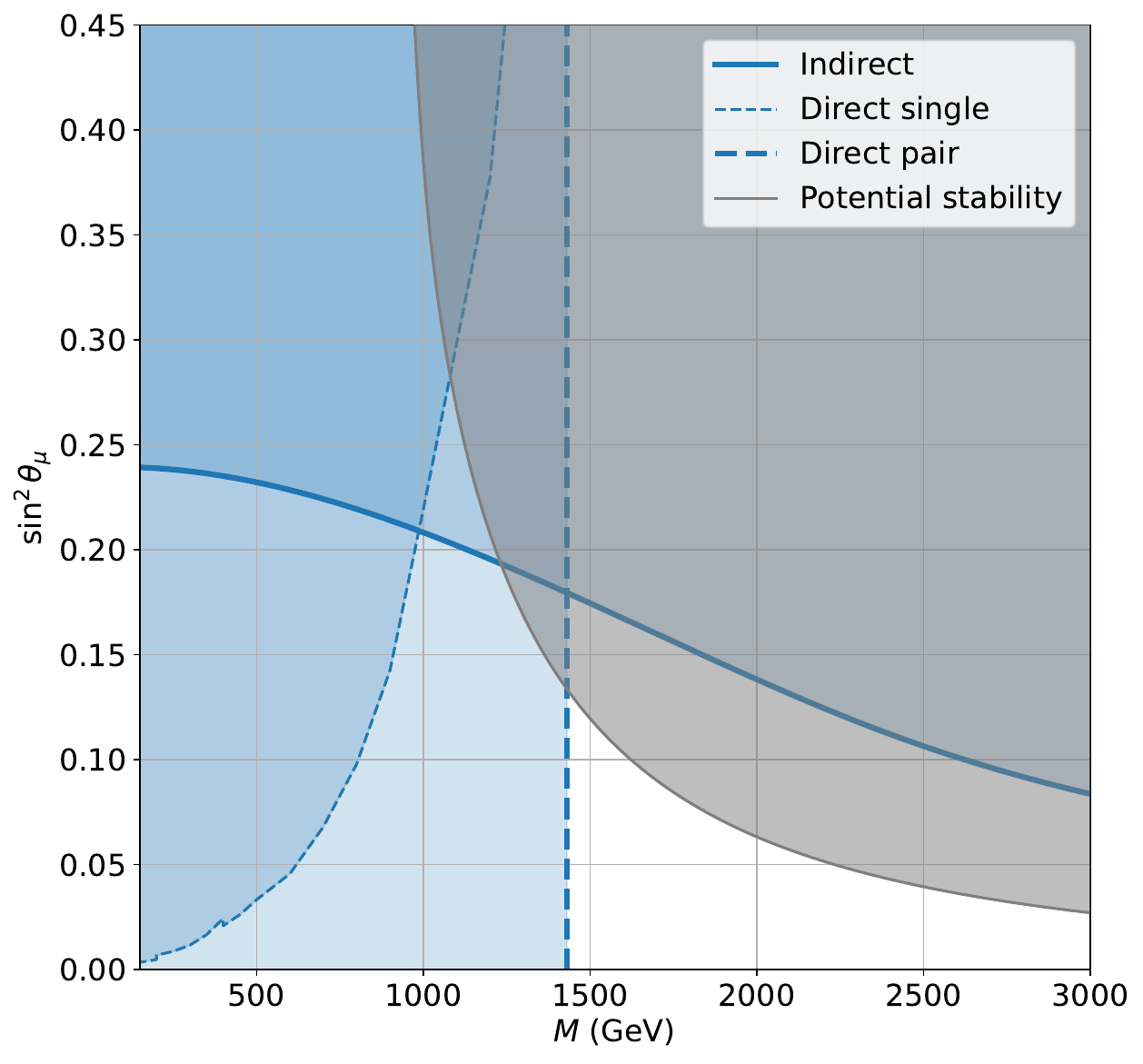}
    \includegraphics[width=0.48\linewidth]{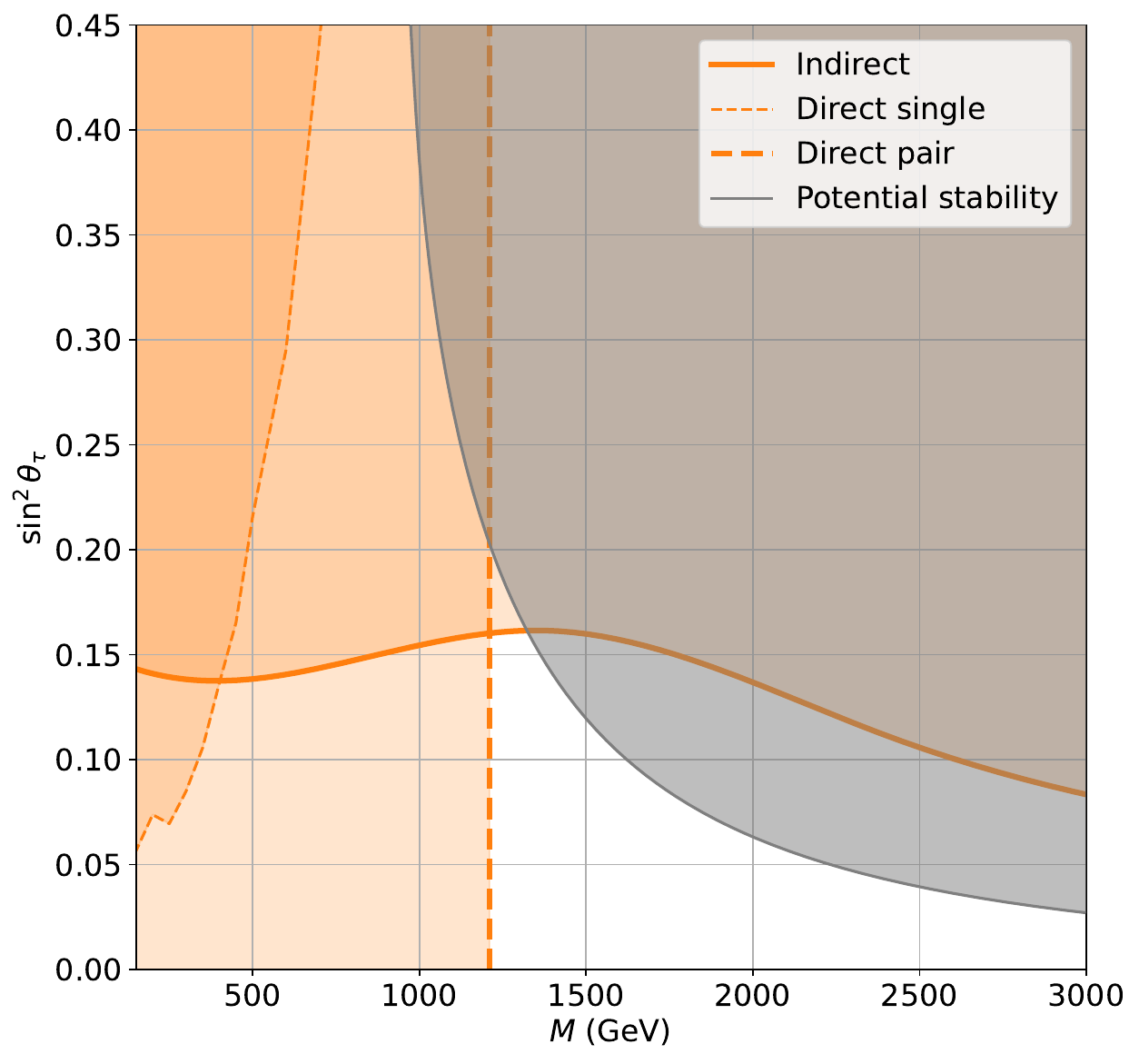}
    \caption{Combination of all exclusion regions at $95\%$ C.L. for the case of mixing with the muon (left) and tau lepton (right). }
    \label{fig:excl_global_mu_tau}
\end{figure}

Once we have explored the constraints on the model arising from different sources, from direct single and pair-production searches to indirect effects from tree-level dimension 6 and 8 operators and one-loop dimension 6 operators, we can combine all these constraints to obtain a global picture of what the current limits on lepton mixing are. We show the combination of all limits, including the theoretical one coming from the stability of the potential, in Figure~\ref{fig:excl_global_e} for mixing with the electron, and Figure~\ref{fig:excl_global_mu_tau} for mixing with the muon (left panel) and tau lepton (right panel). The corresponding limits on the fraction of non-singlet that the charged lepton RH component has are
\begin{equation}
    \sin^2\theta_i \leq 
    \left \{
    \begin{array}{l}
     0.14~[0.20\mbox{, no theoretical constraints}], \mbox{ mixing with $e$}, \\
     0.13~[0.18\mbox{, no theoretical constraints}], \mbox{ mixing with $\mu$}, \\
     0.16, \mbox{ mixing with $\tau$}, \\
    \end{array}
    \right .
\end{equation}
where the results are also given, in square brackets, with the assumption that theoretical constraints are evaded with the addition of heavy new physics. In the case of mixing with the tau lepton theoretical constraints do not change the current limit. 
These results are in striking contrast with the case of mixing with a single VLL. As an example, the equivalent numbers, from EWPO, can be obtained from~\cite{deBlas:2013gla}. The corresponding limits on the mixing squared are $4\times 10^{-4}$, $2.3\times 10^{-3}$ and $1.2\times 10^{-3}$ for the case of electron, muon and tau mixing with $\Delta_1$, respectively, and  $7.8\times 10^{-4}$, $7.8\times 10^{-4}$ and $2.1\times 10^{-3}$ for the case of electron, muon and tau mixing with $\Delta_3$, respectively.

\section{Future colliders\label{sec:future_colliders}}

Future collider experiments will clearly change the picture presented so far, resulting in stronger bounds on (or an observation of) the effects of lepton mixing. We will consider the impact of the planned high-luminosity phase of the LHC (with an integrated luminosity of $3\,\mathrm{ab}^{-1}$) and a possible future $e^+e^-$ collider followed by a high-energy hadron collider. For concreteness, we will consider the case of the future circular collider (FCC) project, starting with the FCC-ee~\cite{FCC:2018byv,FCC:2018evy,FCC:2025lpp}, a circular $e^+e^-$ collider operating at energies from the $Z$ pole up to 365 GeV, to be followed by the FCC-hh~\cite{FCC:2018byv,FCC:2018vvp,FCC:2025lpp}, a $\sqrt{s}=84.6\, \mathrm{TeV}$ proton-proton collider. 

\subsection{HL-LHC}
The high-luminosity phase of the LHC will have two main effects on the limits presented in Figures~\ref{fig:excl_global_e} and \ref{fig:excl_global_mu_tau}: (\emph{i}) the direct searches limit will scale with the increased luminosity; (\emph{ii}) indirect constrains from Higgs physics and EWPO will also become stronger -- \emph{i.e.} improvement in $h\rightarrow\gamma\gamma$, muon Yukawa, W-mass, etc. 

To address point (\emph{i}) we will use the tool \texttt{Collider Reach}~\cite{cReach} which extrapolates limits obtained at a certain $\sqrt{s}$ and integrated luminosity to other points. In ref.~\cite{Guedes:2021oqx} the use of this tool was validated for VLL searches. For pair production, the extrapolated bound is $M\gtrsim2.17,\,2.22,\,1.96\,\mathrm{TeV}$ for the VLL mixing with electron, muon and tau respectively.

To account for the improvements in indirect measures, point (\emph{ii}), we have updated the relevant observables in \texttt{smelli} with the expected precision to be achieved at HL-LHC, and assuming the SM predictions as central values. To this end, we use the projected signal strengths in ref.~\cite{Cepeda:2019klc}, updated with the improvements in ref.~\cite{CMS:2025hfp} for the Higgs observables. Both experimental and theoretical correlations are taken into account. In terms of EWPOs, we consider the improvement in the $W$-boson mass measurement, with an expected uncertainty of $\sim 5$ MeV \cite{ATLAS:2022hsp}. 
The constraints on the Higgs self-coupling are included in the likelihood directly as a bound on $\delta\kappa_\lambda$, 
describing the shift with respect to the 
SM value -- for more details on the computation of $\delta\kappa_\lambda$ in our model see Appendix D. The projected uncertainties to this quantity are taken from ~\cite{deBlas:2025gyz} (see Sections 3.1.1 and 3.1.3).

The resulting improved constraints from the HL-LHC are shown in Fig. \ref{fig:hl_fccee_indirect} with  dotted lines. The expected upper limit on the mixing squared will be, at the end of the HL-LHC phase
\begin{equation}
    \sin^2\theta_i \leq 
    \left \{
    \begin{array}{l}
     0.05~[\mbox{0.08, no theoretical constraints}], \mbox{ mixing with $e$}, \\
     0.05~[\mbox{0.06, no theoretical constraints}], \mbox{ mixing with $\mu$}, \\
     0.04~[\mbox{0.05, no theoretical constraints}], \mbox{ mixing with $\tau$}. \\
    \end{array}
    \right .
\end{equation}
Once again, we give the results with and without taking into account theoretical constraints.
The increased precision of the HL-LHC makes the effect of theoretical constraints less stringent but still dominant. One could have predicted the result for the muon mixing case, due to the expected 1~$\sigma$ uncertainty associated with the muon Yukawa measurement, which should reach 3\%. As it happened in the tau case for current data, Eq. \eqref{eq:kyukawa} translates the muon Yukawa 95\% C.L. bound into
\begin{equation}
    \sin^2\theta_\mu \lesssim 0.06. 
\end{equation}
Note that this bound is not flat in Fig. \ref{fig:hl_fccee_indirect} as other one-loop effects also play a role.

\subsection{FCC-ee and FCC-hh}

\begin{figure}[h!]
\centering
\begin{subfigure}{.6\textwidth}
  \centering
  \includegraphics[width=.8\linewidth]{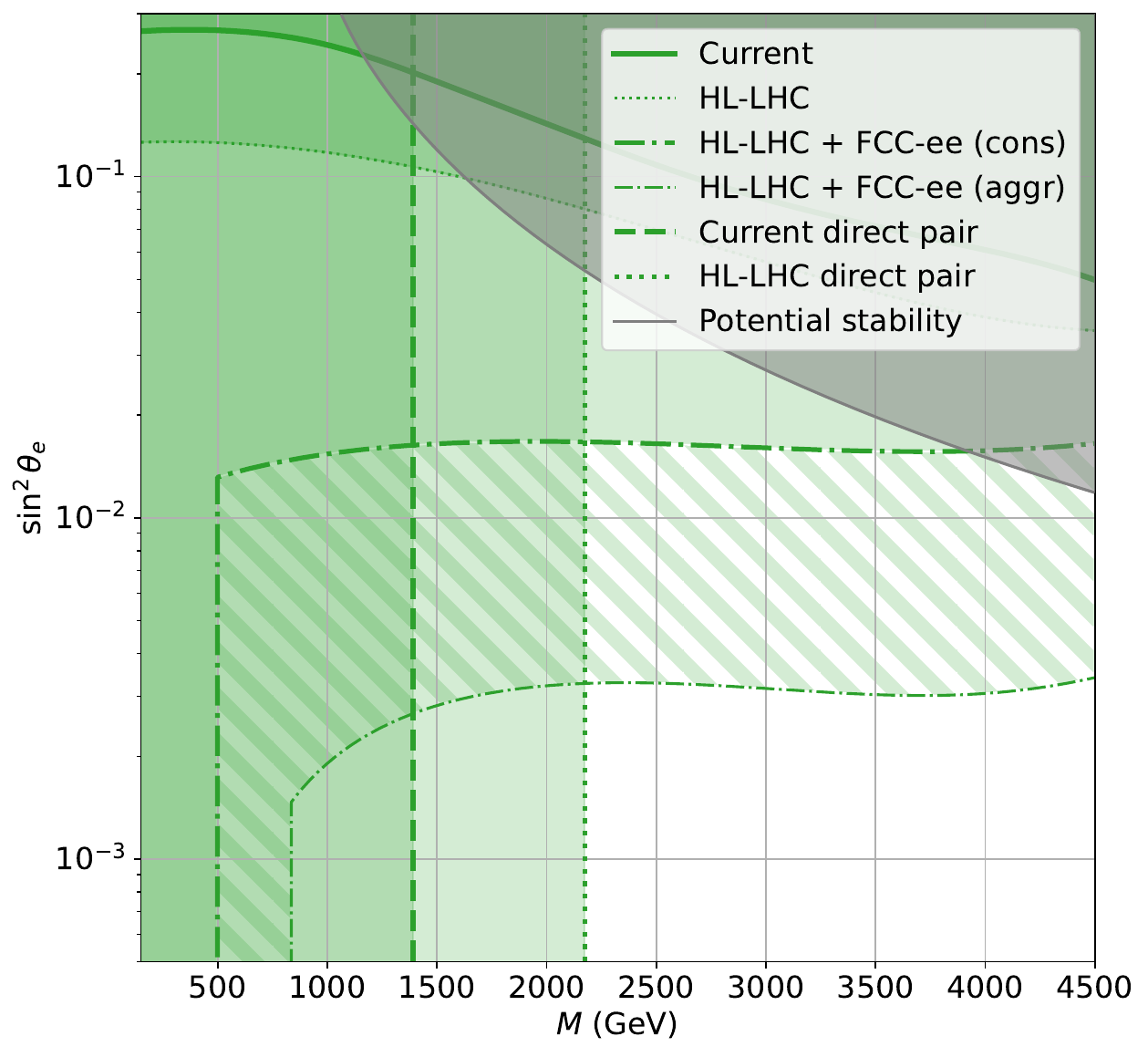}
\end{subfigure}
\\
\begin{subfigure}{\textwidth}
  \centering
  \includegraphics[width=.48\linewidth]{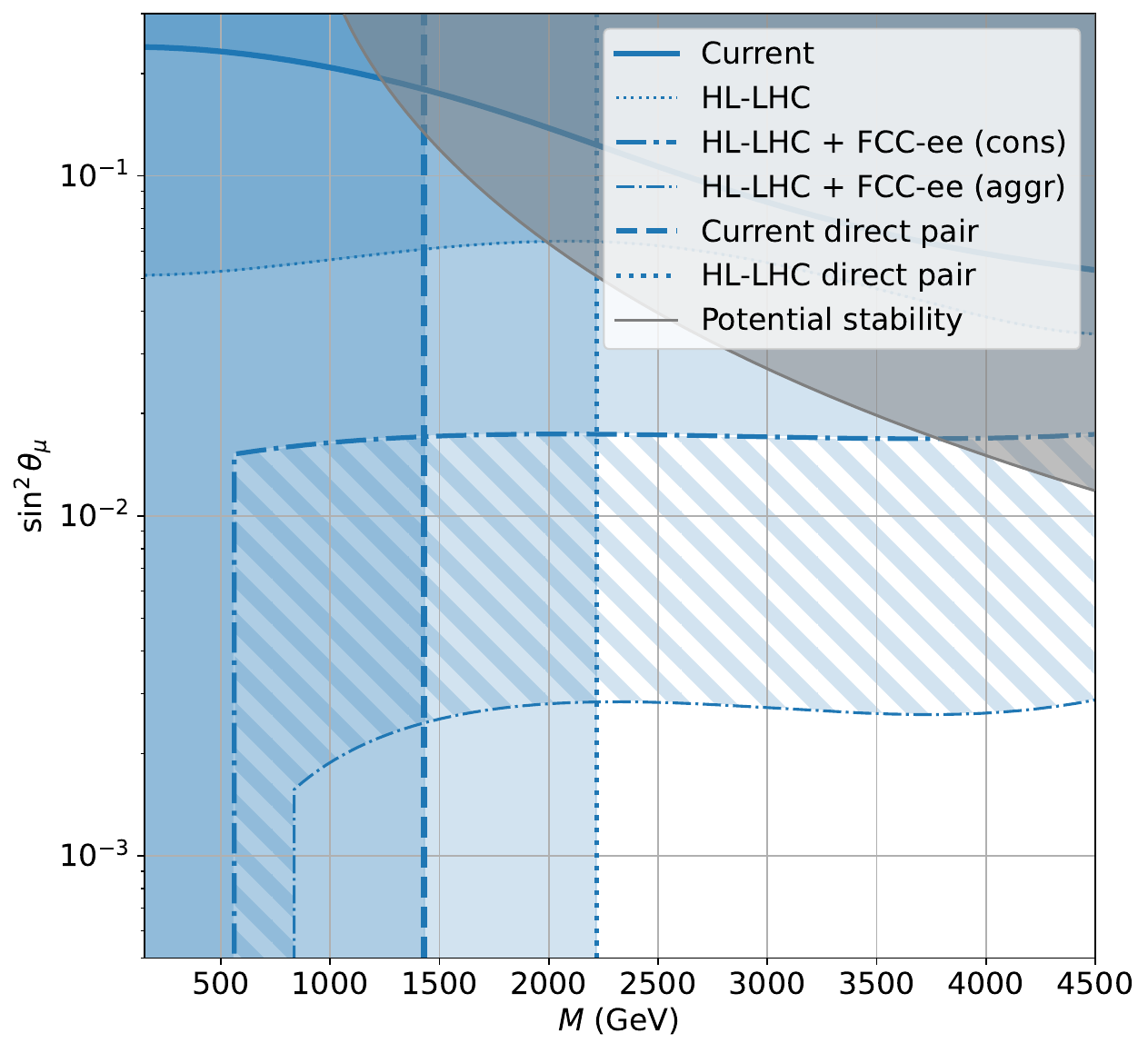}
  \includegraphics[width=.48\linewidth]{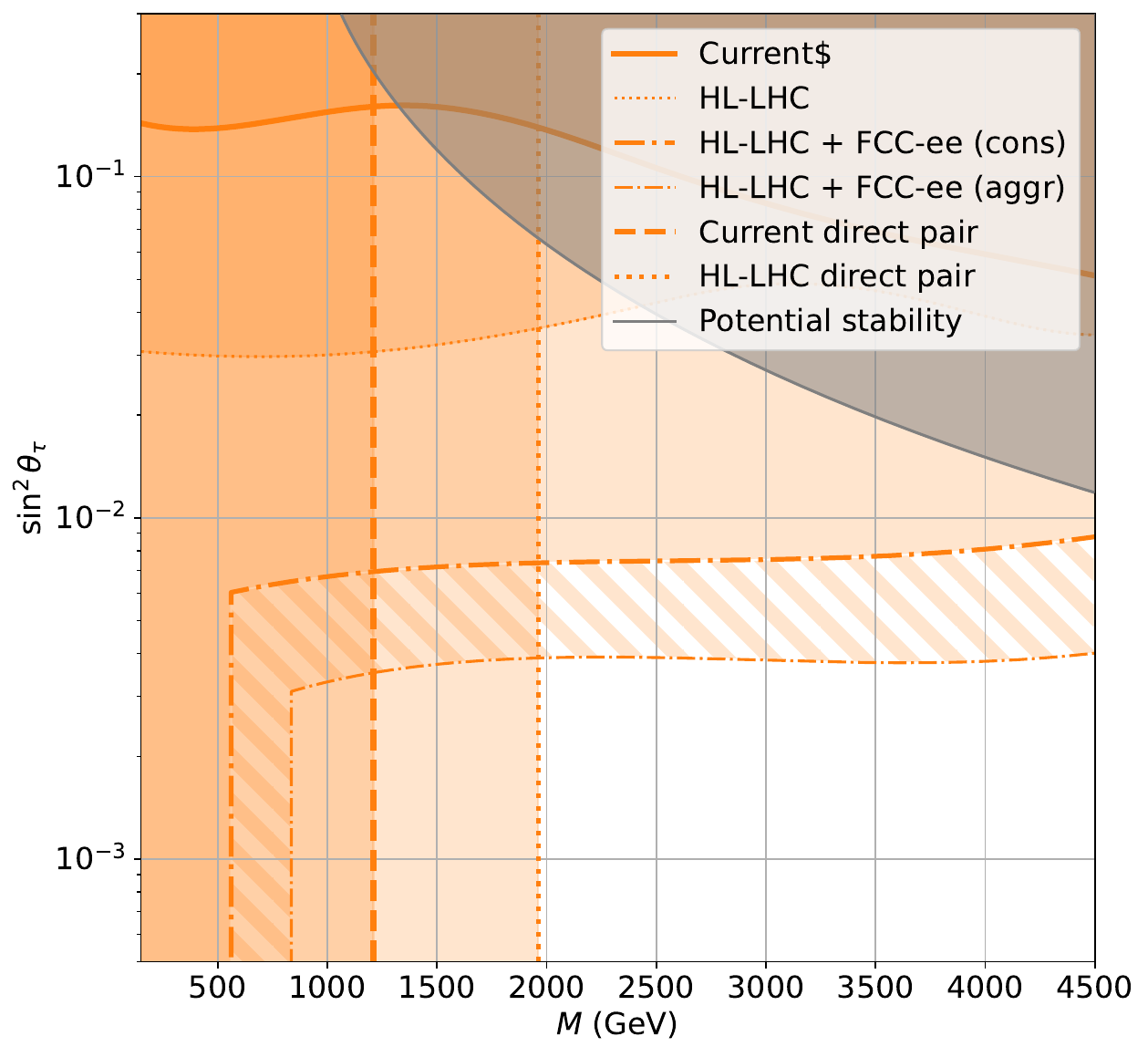}
\end{subfigure}
\caption{Comparison between excluded regions of parameters at 95\% C.L. in log scale using current data (solid lines), HL-LHC projections (dotted lines) from signal strengths assuming trivial correlations (id.), and FCC-ee projections (dashed lines) assuming conservative (thick) and aggressive (thin) uncertainties. We show these for the single mixing with $e$ (green), $\mu$ (blue) and $\tau$ (orange).}
\label{fig:hl_fccee_indirect_log}
\end{figure}

The FCC-ee programme, with a run at $\sqrt{s} \approx M_Z$ that expects to collect $6 \cdot 10^{12}$ Z bosons (the Tera-Z run) will significantly improve the constraints arising from EWPO~\cite{Selvaggi:2025kmd}.
The FCC-ee is also planned to run around the $WW$ threshold, improving the precision of the $W$-bosons properties, and as a Higgs factory at $\sqrt{s}=240$ and $365\,\mathrm{GeV}$, significantly improving the HL-LHC precision in single-Higgs measurements. 
We make use of the observables implementation in \texttt{smelli} and \texttt{flavio} introduced in Ref.~\cite{Allanach:2025wfi}, which define the relevant Higgs physics measurements in FCC-ee at higher energy runs. 
We incorporate these considering the most recent projections for the uncertainties: for the experimental ones, we use the systematic and statistical uncertainties from ~\cite{Selvaggi:2025kmd} (Table 2 for Higgs and Tables 4 - 6 for EWPOs); for the theoretical uncertainties, we present the results for two scenarios, matching the conservative and aggressive assumptions for the projected precision of theory calculations in ~\cite{deBlas:2025gyz} (Tables 3.2 - 3.6). We also take into account and compute the theoretical correlations considering these two scenarios for the uncertainties. 
Finally, the FCC-ee constraints on the Higgs self-coupling from single-Higgs measurements are also taken into account in our fit, including again the bounds reported in~\cite{deBlas:2025gyz}.~\footnote{Note that, in our case, these estimates are expected to be conservative, since our model describes a more restricted scenario compared to the global SMEFT fit used to obtain the $\delta \kappa_\lambda$ limits in that reference.}
\begin{figure}[t!]
\centering
\begin{subfigure}{.6\textwidth}
  \centering
  \includegraphics[width=.8\linewidth]{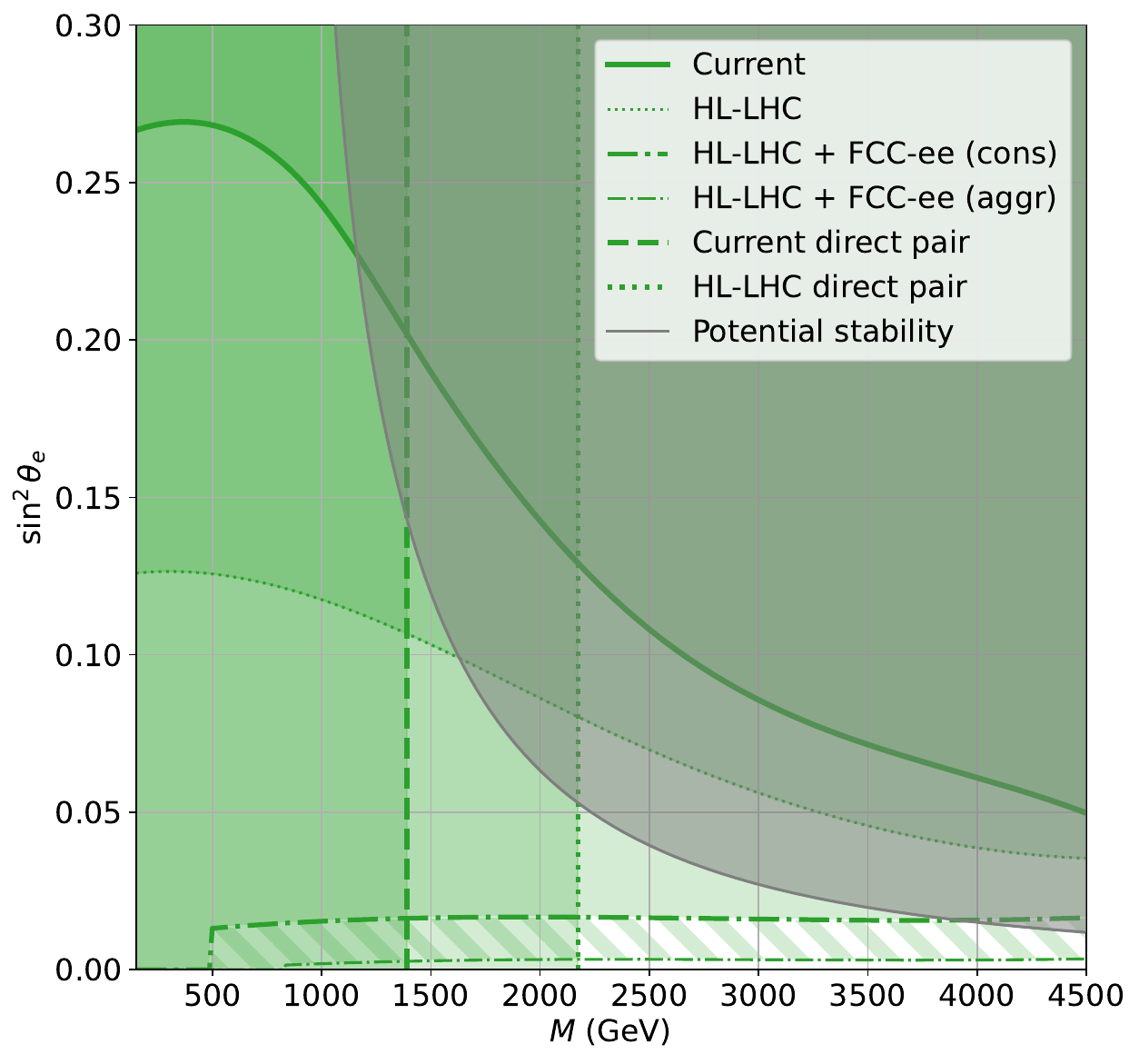}
\end{subfigure}
\\
\begin{subfigure}{\textwidth}
  \centering
  \includegraphics[width=.48\linewidth]{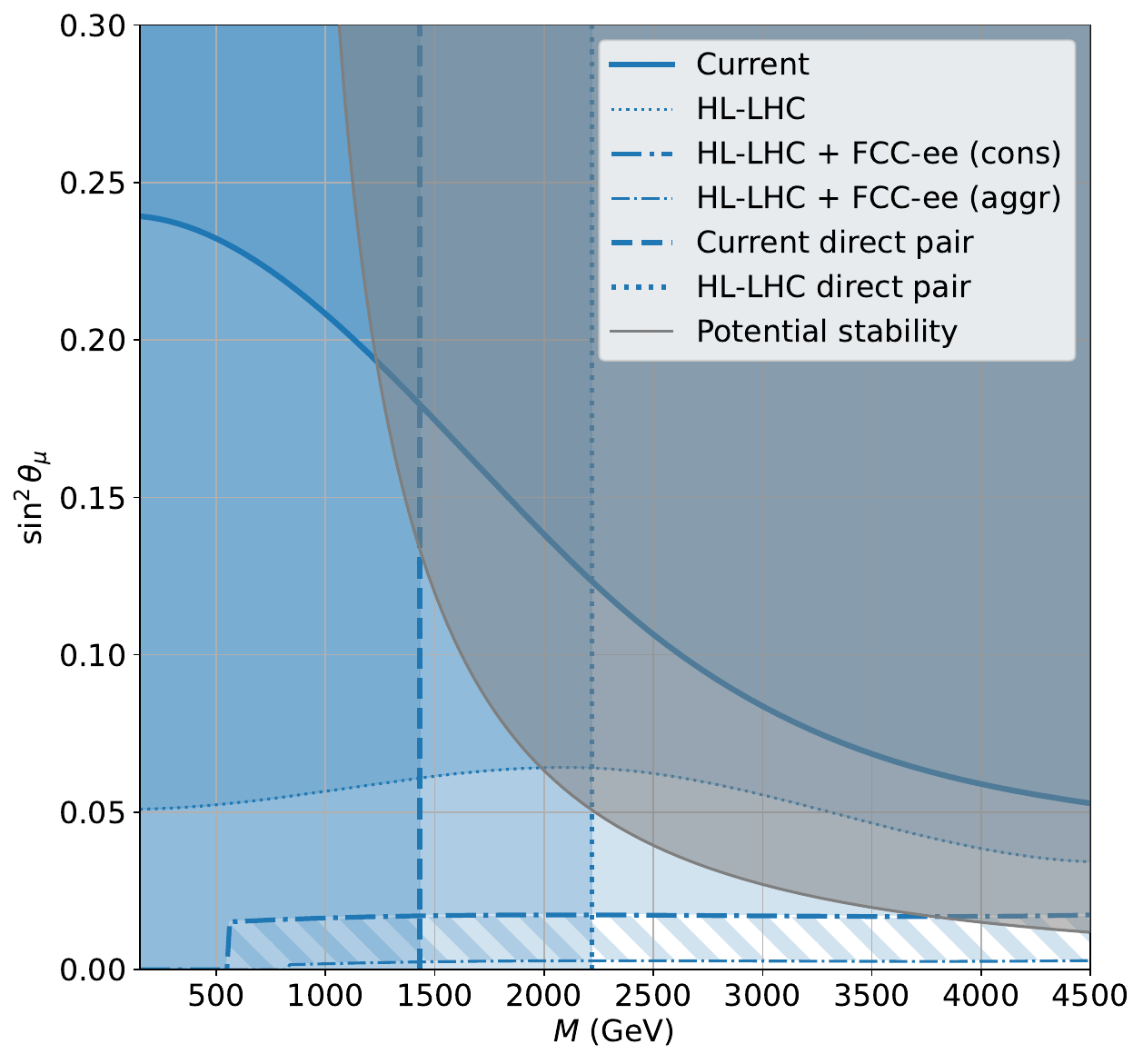}
  \includegraphics[width=.48\linewidth]{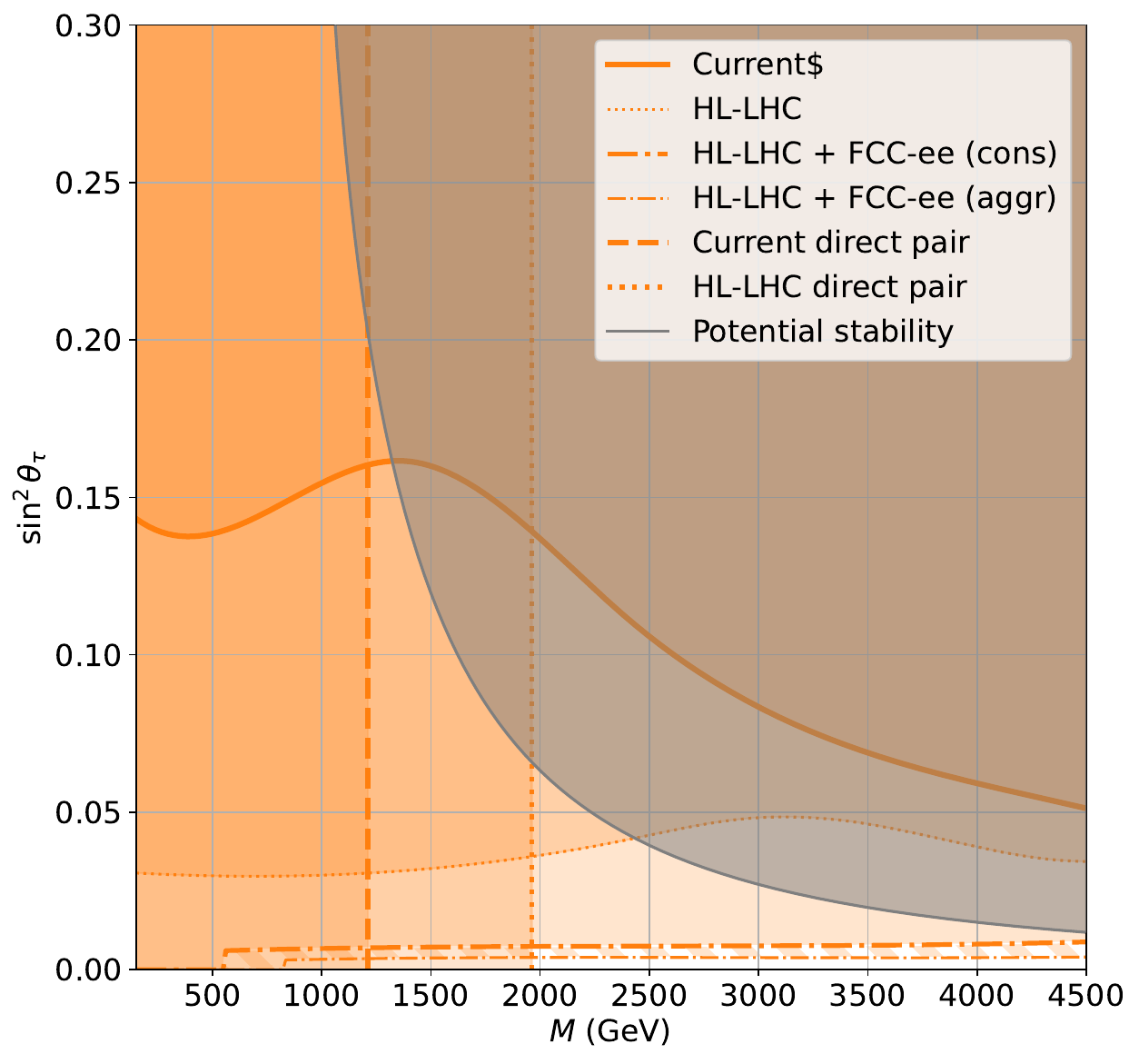}
\end{subfigure}
\caption{Comparison between excluded regions of parameters at 95\% C.L. using current data (solid lines), HL-LHC projections (dotted lines) from signal strengths assuming trivial correlations (id.), and FCC-ee projections (dashed lines) assuming conservative uncertainties. We show these for the single mixing with $e$ (green), $\mu$ (blue) and $\tau$ (orange).}
\label{fig:hl_fccee_indirect}
\end{figure}
The obtained limits are presented in Fig.~\ref{fig:hl_fccee_indirect}, in thick (thin) dash-dotted lines for the conservative (aggressive) assumptions on the theoretical uncertainties. Given the large improvement in the mixing bound in comparison to HL-LHC, we present the same bounds in Fig. \ref{fig:hl_fccee_indirect_log} in logarithmic scale so that extent of the constraints becomes more evident.
The strength of the FCC-ee programme becomes clear upon inspection of Fig.~\ref{fig:hl_fccee_indirect_log}. When taken in combination with direct pair production constraints (from HL-LHC), it is only through the indirect bounds from the FCC-ee that we obtain a stronger constraint than that resulting from theoretical considerations. 
This is in stark contrast with the LHC for which, even in the precision era of the HL-LHC, theoretical considerations play the leading role in constraining the mixing.

The expected upper limit on the mixing squared in the conservative (aggressive) scenario will be, at the end of the FCC-ee, 
\begin{equation}
    \sin^2\theta_i \leq 
    \left \{
    \begin{array}{l}
    17 \ (3) \ \cdot 10^{-3}, \mbox{ mixing with $e$,\text{ cons. (aggr.)}}, \\
    17 \ (3) \ \cdot 10^{-3}, \mbox{ mixing with $\mu$,\text{ cons. (aggr.)}}, \\
    7 \ (4) \ \cdot 10^{-3}, \mbox{ mixing with $\tau$,\text{ cons. (aggr.)}}. \\
    \end{array}
    \right .
    \label{eq:mixing_limits_fccee}
\end{equation}

We note that, while the bounds on mixing with electrons/muons are controlled by EWPO, in the conservative theory scenario mixing with the tau receives stronger constraints from the improved $H\to \tau\tau$ FCC-ee measurements -- see Refs.~\cite{Erdelyi:2025axy,Allwicher:2025mmc} for studies of models with large contributions to the electron Yukawa and how the FCC-ee plays a role there. Assuming theory uncertainties improve to the aggressive case, however, makes this scenario also strongly constrained by the FCC-ee $Z$-pole measurements.

Complementarily to the FCC-ee programme, a future hadron collider would be able to directly probe higher masses of the degrees of freedom of our model. Once again, using the \texttt{Collider Reach} tool~\cite{cReach}, we can extrapolate the mass limits obtained with current data. Note that this is a conservative result as, given the large center of mass energy and the improvements in data analysis, a much better tailored analysis could be developed for our model. We neglect expected improvements in indirect bounds arising at FCC-hh. Considering the FCC-hh with $\sqrt{s}=84.6\,\mathrm{TeV}$ and $\mathcal{L}=30\,\mathrm{ab}^{-1}$~\cite{FCC:2025lpp}, the direct pair production bounds are 10.9, 11.2, 9.7 TeV for mixing with electron, muon and tau respectively.

This expected improvement in direct searches would turn the situation into the current paradigm, described in Section~\ref{sec:combination}, where the combination of theoretical bounds and direct pair production limits are more constraining than indirect precision measurements. At $M\sim10$ TeV, the stability of the potential $\lambda^\prime\lesssim2$ provides a bound on the mixing squared, $\sin^2\theta\lesssim2.4\cdot10^{-3}$, which is stronger than the FCC-ee limit in the aggressive scenario, see Eq.~\eqref{eq:mixing_limits_fccee}. Even when accounting for the expected increased precision of Higgs observables from FCC-hh, the situation does not change, since double Higgs production, the dominant observable for large $\lambda '$, will have an expected precision on the modification of the Higgs self-coupling $\delta\kappa_\lambda\lesssim0.05$~\cite{Selvaggi:2025kmd}; using Eq.~\eqref{eq:kappa_lambda}, this would result in a limit on the mixing squared of $\sin^2\theta\lesssim6.5\cdot10^{-3}$ at 95\%~C.L.. For this bound to be at least as strong as the theoretical constraint $\lambda^\prime\lesssim2$ at $M\sim10$ TeV, the required precision on $\delta\kappa_\lambda$ would be $\delta\kappa_\lambda\lesssim0.005$, an order of magnitude more precise than what FCC-hh is expected to achieve.

\section{Conclusions and outlook \label{sec:conclusions}}

The coupling of the charged leptons to the EW gauge bosons has been experimentally measured with a precision better than \textit{per mille} level and they agree with the SM predictions to the same level of precision.
This seems to prevent a significant mixing of the SM leptons with new, heavy leptons, as this mixing generically induces anomalous couplings of the physical leptons to the EW gauge bosons. We have shown that this intuition is not necessarily correct in the case that the SM leptons can mix with more than one multiplet of heavy leptons. In this case it is possible to have cancellations that suppress these anomalous couplings even in the presence of a significant mixing. These cancellations can be furthermore guaranteed by symmetries, thus making them natural.

We have studied a specific example of such a scenario in the so-called degenerate bidoublet model. The SM leptons mix, with equal strength, to two degenerate lepton doublets with hypercharges $-1/2$, $\Delta_1$, and $-3/2$, $\Delta_3$, respectively. Integrating out the heavy leptons we obtain that the only operator in the SMEFT at dimension 6 that is generated at tree level is suppressed by the charged lepton Yukawa couplings and, in practice, only results in a re-scaling of such a coupling. This results only in a relatively mild constraint in the case of the tau lepton, and it is completely irrelevant for the first two generations. We have carefully considered other sources of constraints on the lepton mixing in the model and have found that direct searches for pair production in combination with the effect of dimension 6 operators generated at one-loop order in Higgs physics and EWPO (and $g-2$ in the case of mixing with the muon) produce the strongest constraint on the allowed mixing. These constraints are nevertheless quite mild. We have found that the RH electron could have a $20\%$ component of non-singlet while the corresponding numbers for the muon and tau lepton are, respectively, $18\%$ and $16\%$. These bounds are so mild that theoretical considerations become, in combination with direct searches, more stringent than indirect precision constraints, leading to an upper limit on the allowed non-singlet component of $14\%$ for the $e$ and $13\%$ for the muon (mixing with $\tau$ is not more strongly constrained by theoretical considerations).

Future experimental results will significantly strengthen these limits or, alternatively, produce evidence of the mixing effects. The improvement in the Higgs measurements at the HL-LHC, together with a higher reach in direct searchers, will constrain the amount of non-singlet in the RH electron, muon and tau to $8\%$, $6\%$ and $5\%$, respectively. The improved limit on the mass of the heavy fermions from direct searches makes theoretical considerations again more constraining, leading to $5\%$ for the electron and muon, and $4\%$ for the tau. The FCC-ee will improve these limits further, thanks to the precision obtained in EWPO and Higgs measurements, taking the limits down to   
$0.3-1.7\%$, $0.3-1.7\%$ and $0.4-0.7\%$, depending on the assumptions on the size of theory uncertainties for EWPO. 
The precision is high enough that these limits are more stringent than the ones derived from theoretical considerations.
Finally, the FCC-hh will significantly extend the reach in direct searches to $M\sim 10$ TeV, at which point theory bounds constrain $\sin^2{\theta}\lesssim 2.4\cdot 10^{-3}$.

The fact that current limits on the mixing are so mild and the rich phenomenology of the degenerate bidoublet model provides exciting prospects of finding new physics in the near future. Dedicated searches, including single production of charged VLLs, that we plan to take on future work, will be an excellent opportunity to discover new particles or further carve the allowed parameter space. Such a discovery could be an excellent opportunity to composite Higgs models that naturally incorporate the custodial protection that is realized in our model~\cite{delAguila:2010vg,delAguila:2010es,Carmona:2013cq}.

\acknowledgments

We would like to thank F. del Águila, M. Chala and J. Fuentes-Martín for useful discussions. This work has been partially supported by the Ministry of Science and Innovation and SRA (10.13039/501100011033) and ERDF under grants PID2022-139466NB-C21 and EUR2024-153549, by the Junta
de Andaluc\'ia grant FQM 101, by Consejer\'{\i}a de Universidad, Investigaci\'on e Innovaci\'on, Gobierno de España and Uni\'on
Europea – NextGenerationEU under grant $\mathrm{AST}22\_6.5$ and by the Deutsche Forschungsgemeinschaft (DFG, German Research Foundation) under grant 491245950 and under Germany’s Excellence Strategy — EXC 2121 “Quantum Universe” — 390833306. RSL is further supported by an
FPU grant (FPU23/03050) from Consejería de Universidad, Investigación e Innovación, Gobierno
de España. This work has been partially funded by the Eric \& Wendy Schmidt Fund for Strategic Innovation through the CERN Next Generation Triggers project under grant agreement number SIF-2023-004.

\appendix

\section{\label{appendix:oneloop:matching} One loop matching in the Warsaw basis}

The complete result for the one-loop matching in the Warsaw basis is very long and is provided as an ancillary file in the arXiv submission of this work. We list here the results that are relevant for the discussion of indirect constraints  in Section~\ref{sec:one-loop-dim6} and also for di-Higgs production and theoretical limits based on the stability of the potential. In particular, we will write only the terms that are enhanced by powers of the $\lambda^\prime_i$ coupling, except whenever they are suppressed by lepton Yukawa couplings~\cite{Carmona:2021xtq,Fuentes-Martin:2022jrf,Guedes:2023azv,Guedes:2024vuf}: 

\begin{align}
16\pi^2 \wc_{\phi} &= \sum_{ijk} \frac{8}{3} \frac{|\lambda^\prime_i|^2 |\lambda^\prime_j|^2|\lambda^\prime_k|^2}{M^2} + \ldots,
\label{eq:alphaOH}
\\
16\pi^2 \wc_{\phi B}&= \frac{2g_1^2}{3}\sum_k \frac{ |\lambda_k^\prime|^2}{M^2} +\ldots,
\label{eq:alphaOHB}
\\
16\pi^2 \wc_{\phi\Box} &= \left[\frac{g_1^2}{9}+\frac{g_2^2}{3}\right] \sum_k\frac{ |\lambda^\prime_k|^2}{M^2 }
-\frac{1}{3} \sum_{kl} \frac{|\lambda^\prime_k|^2|\lambda^\prime_l|^2}{M^2}
+\ldots,
\label{eq:alphaOHBox}
\\
16\pi^2 \wc_{\phi D}&= 
\frac{4g_1^2}{9}\sum_k \frac{ |\lambda_k^\prime|^2}{M^2} +\ldots,
\label{eq:alphaOHD}
\\
16\pi^2 \big(\wc_{\phi e}\big)_{ij} &=
-\frac{2g_1^2}{9}  \sum_k \frac{|\lambda^\prime_k|^2}{M^2}\delta_{ij}
+g_1^2 \left[\frac{19}{9} +3 \log\left(\frac{\mu^2}{M^2}\right) \right] \frac{\lambda^{\prime\,\ast}_i \lambda^\prime_j}{M^2} + \ldots,
\label{eq:alphaOHe}
\\
16\pi^2 \big(\wc_{\phi \psi}^{(1)}\big)_{ij} &= 
\frac{2 g_1^2}{9} Y_\psi  \sum_k \frac{|\lambda^\prime_k|^2}{M^2}\delta_{ij} + \ldots,
\label{eq:alphaOHpsi1}
\\
16\pi^2 \big(\wc_{\phi l}^{(3)}\big)_{ij} &= 
16\pi^2 \big(\wc_{\phi q}^{(3)}\big)_{ij} =
\frac{g_2^2}{9}  \sum_k \frac{|\lambda^\prime_k|^2}{M^2}\delta_{ij} + \ldots,
\label{eq:alphaOHpsi3}
\end{align}
where $\wc_{\phi\psi}^{(1)}$ stands for $\wc_{\phi l}^{(1)}$, $\wc_{\phi q}^{(1)}$, $\wc_{\phi u}$ and $\wc_{\phi d}$, for $\psi=l,q,u,d$, respectively, $Y_\psi=-1/2,~\!1/6,$ $2/3,~\!-1/3$ for $\psi=l,q,u,d$, respectively, is the hypercharge of the different fermions and the dots denote terms that are either independent of $\lambda^\prime$ or are suppressed by lepton Yukawa couplings. There are some other operators generated at this order that are proportional to $\lambda^\prime$ and not suppressed by lepton Yukawa couplings but they are not relevant for the global fit.

Two other WCs that are relevant, despite the fact that they are suppressed by light Yukawa couplings, are
\begin{align}
16 \pi^2 \big(\wc_{eW}\big)_{ij} &=
-y^l_i\frac{g_2}{8}  \frac{\lambda^{\prime\,\ast}_i \lambda^\prime_j}{M^2},
\label{eq:alphaOeW}
\\
16 \pi^2 \big(\wc_{eB}\big)_{ij} &=
-y^l_i\frac{13 g_1}{24}  \frac{\lambda^{\prime\,\ast}_i \lambda^\prime_j}{M^2}.
\label{eq:alphaOeB}
\end{align}

Several features of these WCs are worth discussing in detail. The WCs in Eqs. \eqref{eq:alphaOHB} and \eqref{eq:alphaOHBox} enter Higgs physics observables while the ones in Eqs.~\eqref{eq:alphaOHD}-\eqref{eq:alphaOHpsi3} enter EWPO. The latter ones are all $\sim \lambda^{\prime\,2}/M^2$. Thus, a global fit using only EWPO would result in a limit on $(m^\prime/M)^2$ that is independent of $M$. 
Adding Higgs physics observables to the fit, we have $\wc_{\phi\Box}$ that has an extra $ \sim \lambda^{\prime\,4}/M^2$ contribution that breaks this degeneracy and results in a mass-dependent limit. A detailed analysis shows that despite the fact that EWPO have a much higher precision than Higgs observables, the contribution to $H \to \gamma \gamma$ has to compete with a SM one-loop effect, while a similarly sized contribution has to compete with a tree-level one in the SM for EWPO. In the end Higgs physics observables are more competitive in our model and they dominate the fit. In the case of the electron, while subdominant, EWPO play a non-negligible role, making the final result more constraining than for the muon.

Another feature that might seem confusing at first sight is the presence of a logarithm in Eq.~\eqref{eq:alphaOHe}, given that in the limit of vanishing lepton Yukawa couplings there are no tree level contributions to the matching and, therefore, no logarithmic dependence should appear in the one-loop matching results. The reason is that the results above are given in terms of the $\overline{\mathrm{MS}}$-renormalized parameters of the model at the renormalization scale $\mu$. If we express the results in terms of the parameters at a fixed renormalization scale the logarithms of the renormalization scale indeed disappear. Let us discuss this in more detail in the example of $\wc_{\phi e}$ in a simplified scenario where we neglect all SM Yukawa couplings and assume mixing with a single generation. It should be noted first that the condition that the two doublets have the same mass and identical coupling to the SM leptons is not radiatively stable, due to the different hypercharges of the two doublets. Indeed, denoting $M_{1,3}$ and $\lambda^\prime_{1,3}$ the mass and coupling of $\Delta_1$ and $\Delta_3$, respectively, we find,
\begin{equation}
\wc_{\phi e}= \frac{1}{2}\left(\frac{|\lambda^\prime_1|^2}{M_1^2}-\frac{|\lambda^\prime_3|^2}{M_3^2} \right) +\frac{1}{16\pi^2}\frac{3 g_1^2 |\lambda^\prime_1|^2}{M_1^2} \log\left(\frac{\mu^2}{M^2}\right) + \ldots,
\end{equation}
where all the parameters appearing in this expression are renormalized parameters evaluated at the renormalization scale $\mu$. We can compute the RGEs of the full model to express them in terms of the parameters at a fixed scale $\mu_0$. We obtain
\begin{align}
\label{eq:uvrges}
\dot{M}_1 &= [-3 g_1^2 -9 g_2^2 + |\lambda^{\prime}_{1}|^2 ]\frac{M_1}{2}, \\ 
\dot{M}_3 &= [-27 g_1^2 -9 g_2^2 + |\lambda^{\prime}_{3}|^2 ]\frac{M_1}{2}, \\    
\dot{\lambda^\prime_{1}} &= \frac{\lambda^\prime_1}{4} [-15g_1^2 -9 g_2^2 + 10 |\lambda^\prime_1|^2
+ 16 |\lambda^\prime_3|^2], \label{eq:lam1pdot}\\
\dot{\lambda^\prime_{3}} &= \frac{\lambda^\prime_3}{4} [-39g_1^2 -9 g_2^2 + 16 |\lambda^\prime_1|^2
+ 10 |\lambda^\prime_3|^2], \label{eq:lam3pdot}
\end{align}
so that we can write, consistently at fixed one-loop order
\begin{equation}
\wc_{\phi e}(\mu)= \frac{1}{2}\left(\frac{|\lambda^\prime_1|^2}{M_1^2}-\frac{|\lambda^\prime_3|^2}{M_3^2} \right)_{\mu=\mu_0} +\frac{1}{16\pi^2}\frac{3 g_1^2 |\lambda^\prime_1|^2}{M_1^2} \log\left(\frac{\mu_0^2}{M^2}\right) + \ldots.
\end{equation}
Note that, as explicitly stated on the LHS of the previous equation, we have not just evaluated the WC at a fixed renormalization scale $\mu_0$ but we have expressed the previous result in terms of renormalized parameters at that scale. It turns out that in this case the $\mu$ dependence completely disappears, as is expected from the fact that no tree-level dimension-6 contribution survives in the Yukawa-less limit we are considering.

\section{Constraints from dipole moments\label{app:gminus2}}

The contribution to dipole moments is particularly simple in our model, given that all relevant contributions are, either one-loop generated or have an extra suppression by the light lepton masses. Ignoring the latter, the only one-loop contribution to the anomalous magnetic moment of the charged leptons come from the dipole operator in the LEFT Lagrangian~\cite{Aebischer:2021uvt}
\begin{equation}
    \Delta a_l = \frac{4 m_l}{e q_e} ~\Re[ \big(L_{e\gamma}\big)_{ll} ], 
\end{equation}
where $q_e=-1$ is the charge of the lepton and we have changed the sign of the QED coupling constant, $e$, given the different conventions in the covariant derivative between our work and~\cite{Aebischer:2021uvt}.\footnote{EDMs do not introduce any further constraint. In the case of mixing with a single generation the phase of $\lambda^\prime$ can be reabsorbed in the fields $\Delta_{1,3}$ and it is therefore non-physical. Even in the case of mixing with several generations this phase will be suppressed by the product of different $\lambda^\prime_{i,j}$, which is very strongly constrained by LFV observables, and furthermore, we have checked, following~\cite{Panico:2018hal} that no EDM is generated in our model up to the two loop order.}
The WC $L_{e\gamma}$ is defined in the LEFT as 
\begin{equation}
\mathcal{L}_{\textrm{LEFT}} = \big(L_{e\gamma}\big)_{ij} \bar{e}_{L\,i} \sigma^{\mu \nu} e_{R\,j} F_{\mu \nu} + \ldots,
\end{equation}
with $F_{\mu\nu}$ the electromagnetic field strength tensor. The WC can be obtained in terms of the WCs in the SMEFT by using the tree-level matching between them~\cite{Jenkins_2018} (other contributions are two loops or higher),
\begin{equation}
    \big(L_{e\gamma}\big)_{ij} = v \left[ - s_W \big(C_{eW}\big)_{ij}
    +c_W \big(C_{eB}\big)_{ij}\right],
\end{equation}
where the sine and cosine of the weak angle are given by the expressions $e=g_2 s_W= g_1 c_W$ and, again, the small differences with respect to the original result in~\cite{Jenkins_2018} are due to different notation ($v=v_T/\sqrt{2}$) and to the fact that the input parameters receive no corrections at this order in our model.

Using the expressions for the SMEFT WCs in terms of $\lambda^\prime$ and $M$, Eqs.~\eqref{eq:alphaOeW} and~\eqref{eq:alphaOeB} we obtain
\begin{equation}
\Delta a_l = \frac{5}{96 \pi^2} \frac{m_l^2}{v^2} 2 \left(\frac{m^\prime_l}{M}\right)^2.
\end{equation}

The best current numbers for the electron and muon anomalous magnetic couplings are
\begin{align}
a^{\mathrm{exp}}_e&=1 159 652 180.73(28) \times 10^{-12}, \quad \text{\cite{Hanneke_2008}},
\\
a^{\mathrm{exp}}_\mu &=116 592 071.5(14.5) \times 10^{-11}, \quad \text{\cite{Aliberti:2025beg}}.
\end{align}
Taking the experimental error as the maximum allowed deviation in our model we get, at $95\%$ C.L.,
\begin{align}
2\left(\frac{m_e^\prime}{M}\right)^2 &\leq 1285, 
\label{eq:mdmebound}
\\
2\left(\frac{m_\mu^\prime}{M}\right)^2 &\leq 0.15. 
\label{eq:mdmmubound}
\end{align}
The magnetic dipole moment constraints are therefore only relevant for the muon case. Importantly, the bound of \eqref{eq:mdmmubound} is a conservative one, assuming that the central experimental value is fully accounted for by the SM such that the new physics contribution is bounded by the experimental precision. However, the current SM prediction is slightly below the experimental value (less than $1\sigma$ discrepancy) and a small positive correction from our model could improve a fit to data.

\section{Lepton flavor violating constraints \label{app:LFV}}

The Low-Energy Effective Lagrangian relevant for the description of the processes $l\to 3 l^\prime$ and $l \to l^\prime \gamma$ reads~\cite{Crivellin_2017}: 
\begin{align}
\mathcal{L}_{\rm eff}&=\mathcal{L}_{\rm QED} + \mathcal{L}_{\rm QCD} 
\nonumber \\[5pt]
&+\frac{1}{\Lambda^2}\bigg\{C_L^DO_L^D 
+ \sum\limits_{f = q,\ell } {\left( 
      {C_{ff}^{V\;LL}O_{ff}^{V\;LL} + C_{ff}^{V\;LR}O_{ff}^{V\;LR} 
     + C_{ff}^{S\;LL}O_{ff}^{S\;LL}} \right)}\nonumber \\
& \qquad
+ \sum\limits_{h = q,\tau } {\left( 
       {C_{hh}^{T\;LL}O_{hh}^{T\;LL} + C_{hh}^{S\;LR\;}O_{hh}^{S\;LR\;}} \right)}+
C_{gg}^LO_{gg}^L  + L \leftrightarrow R\bigg\} + \mbox{h.c.},
\label{Leff}
\end{align}
where the different operators are defined by
\begin{align}
\label{eq:magnetic}
O_L^{D} &= e\, m_\mu\left( \bar e{\sigma ^{\mu \nu }}{P_L}\mu\right) {F_{\mu \nu }},
\\[3pt]
O_{ff}^{V\;LL} &= \left(\bar e{\gamma ^\mu }{P_L}\mu\right) 
\left( \bar f{\gamma _\mu }{P_L}f\right),
\\[3pt]
O_{ff}^{V\;LR} &= \left(\bar e{\gamma ^\mu }{P_L}\mu\right) 
\left( \bar f{\gamma _\mu }{P_R}f\right),
\\[3pt]
O_{ff}^{S\;LL} &= \left(\bar e{P_L}\mu\right) \left( \bar f{P_L}f\right),
\\[3pt]
O_{hh}^{S\;LR} &= \left(\bar e{P_L}\mu\right) \left( \bar h{P_R}h\right),
\\[3pt]
O_{hh}^{T\;LL} &= \left(\bar e{\sigma _{\mu \nu }}{P_L}\mu\right) 
\left( \bar h{\sigma ^{\mu \nu }}{P_L}h\right),
\\[3pt]
O_{gg}^L &= \alpha_s\, m_\mu G_F
\left( \bar e P_L\mu  \right)G^a_{\mu \nu }G^{a~\!\mu \nu },
\label{eq:Ogg}
\end{align}
with $P_{LR}=\left(1\mp \gamma^5\right)/2$ are the chirality projectors and $f$ runs over all SM fermions except for the top quark, while $h = \{u, d, s, c, b, \tau\}$. In these equations we have particularized to the $\mu \to e$ transition but they can be trivially generalized to other LFV transitions.

The relevant observables are, again for the particular $\mu \to e$ case,~\cite{Crivellin_2017}
\begin{align}
\label{muegBR}
{\rm{Br}}\left( {\mu  \to e\gamma } \right)  &=  
\frac{\alpha_e m_\mu ^5}{{\Lambda^4 \Gamma _\mu }}\left( 
{{{\left| {C^{D}_L} \right|}^2} 
+ {{\left| {C^{D}_R} \right|}^2}} \right)\,,
\\
{\rm Br}(\mu  \to 3e) &=
\frac{\alpha_e^2 m_\mu^5 }{12 \pi \Lambda^4 \Gamma _\mu}
\left(\left|C^{D}_{L}\right|^2+\left|C^{D}_{R}\right|^2\right)
\left(8 \log\left[\frac{m_\mu}{m_e}\right]-11\right)+ X_\gamma
\nonumber \\ 
&+\frac{m_\mu^5}{3  (16\pi)^3 \Lambda^4 \Gamma _\mu}
\bigg(\, \left|C_{ee}^{S\;LL}\right|^2 + 
16 \left|C_{ee}^{V\;LL}\right|^2 + 8 \left|C_{ee}^{V\;LR}\right|^2  \nonumber\\
& \phantom{+\frac{m_\mu^5}{3  (16\pi)^3 \Lambda^4 \Gamma _\mu}} 
+\
\left|C^{S\;RR}_{ee}\right|^2 +
16 \left|C_{ee}^{V\;RR}\right|^2 + 8 \left|C_{ee}^{V\;RL}\right|^2\bigg) \, ,
\label{br:meee}
\end{align} 
where $\Gamma _\mu$ is the decay width of the muon, the relevant scale of the processes
is $\mu=m_\mu$, and the interference term with the dipole operator is given by
\begin{align}
X_\gamma ^{}{\rm{ }} =
-\frac{\alpha_e m_\mu^5 }{3 (4\pi)^2 \Lambda^4 \Gamma _\mu}
  (\Re[C^{D}_{L}
  \left(C_{ee}^{V\;RL}+2C_{ee}^{V\;RR}\right)^*]+\Re[C^{D}_{R}
  \left(2C_{ee}^{V\;LL}+C_{ee}^{V\;LR}\right)^*])\, .
\end{align}
For $\mu - e$ conversion it is better to work at the nucleon level, and so we need to change the Lagrangian defined in \eqref{Leff} to contain proton and neutron fields. Doing so results in a shift of the WC of the gluonic operator 
\begin{align}
C_{gg}^L &\to \tilde{C}_{gg}^L = C_{gg}^L - \frac{1}{12 \pi}
\sum_{q=c,b} \frac{C_{qq}^{S\;LL}+C_{qq}^{S\;LR}}{G_F\, m_\mu m_q}
\label{anomaly}
\end{align}
with an analogous equation for $C_{gg}^R$. Then the transition $\Gamma_{\mu\to e}^N = \Gamma(\mu^- N \to e^- N)$ reads
\begin{align}
\Gamma_{\mu\to e}^N &= \frac{m_{\mu}^{5}}{4\Lambda^4}
\left| e\,  C^{D}_{L} \; D_N + 
4\left(
G_F m_\mu m_p \tilde{C}_{(p)}^{SL}  S^{(p)}_N
+   \tilde{C}_{(p)}^{VR} \; V^{(p)}_N
+ p \to n \right) \right|^2 +L\leftrightarrow R,
\label{Gconv}
\end{align}
where 
\begin{align}
\tilde{C}_{(p/n)}^{VR} &= \sum_{q=u,d, s} 
\left(C_{qq}^{V\;RL}+C_{qq}^{V\;RR}\right) \; f^{(q)}_{Vp/n} \, , 
\label{tildeCVR} \\
\tilde{C}_{(p/n)}^{SL} &=  \sum_{q=u,d,s} 
\frac{\left(C_{qq}^{S\;LL}+C_{qq}^{S\;LR}\right)}{m_\mu m_q G_F} 
\; f^{(q)}_{Sp/n}  \ + \ \tilde{C}_{gg}^L \, f_{Gp/n}, 
\label{tildeCSL}
\end{align}
are the effective couplings and $p$ and $n$ denote the proton and the neutron respectively.

At tree level we can obtain the four fermions operators that appear in~\eqref{Leff} by integrating out the Higgs field from the Higgs part of the SM Lagrangian through the equations of motion. By doing so we find that their Wilson coefficients are proportional to the non-diagonal part of the leptons Yukawa couplings induced by the tree level operator $\Op_{e\phi}$ as defined in Eq.~\eqref{eq:yukawa_physical_dim6}. In particular, focusing on $\mu \to e$ processes we have
\begin{align}
\label{eq:Cee}
    C_{ee}^{S\;LL} &= -2C_{ee}^{V\;LR} = \frac{1}{2}\frac{Y_{\mu e}Y_e}{m_H^2},
\\[3pt]
\label{eq:Cqq}
C_{qq}^{S\;LL} &= C_{qq}^{S\;LR} = \frac{1}{4}\frac{Y_{\mu e}Y_q}{m_H^2},
\end{align}
where we have taken the Yukawa couplings to be real and $q={u, d, s, c, b}$. Other WCs are proportional to $Y_{e\mu}\propto m_e/v \ll Y_{\mu e}\propto m_\mu/v $ and have been disregarded.

At one loop the operators in~\eqref{Leff} are induced by the ones we computed in the SMEFT. In particular, the relevant SMEFT operators are (we suppress flavor and gauge indices)
\begin{align}
\Op_{ee} &= (\bar{e} \gamma^{\mu}  e)(\bar{e} \gamma_{\mu}  e),
\nonumber \\
\Op_{le} &= (\bar{e} \gamma^{\mu}  e)(\bar{l} \gamma_{\mu}  l),
\nonumber \\
\Op_{\phi e} &= (\phi^\dagger \ii \overset{\leftrightarrow}{D}_\mu \phi)(\bar{e} \gamma^{\mu} e),
\nonumber \\
\Op_{ed} &= (\bar{d} \gamma^{\mu} d)(\bar{e} \gamma_{\mu}  e),
\label{eq:SMEFT_LFV} \\
\Op_{eu} &= (\bar{u} \gamma^{\mu} u)(\bar{e} \gamma_{\mu} e),
\nonumber \\
\Op_{qe} &= (\bar{q} \gamma^{\mu}q)(\bar{e} \gamma_{\mu} e),
\nonumber \\
\Op_{eW} &= (\bar{l} \sigma^{\mu \nu} e) \sigma^I \phi W^I_{\mu\nu},
\nonumber \\
\Op_{eB} &= (\bar{l} \sigma^{\mu \nu} e)  \phi B_{\mu\nu}. \nonumber
\end{align}
We have matched the WCs of these operators, as obtained in our one-loop matching, to the ones in the LEFT basis, Eqs. \eqref{eq:magnetic} - \eqref{eq:Ogg} using~\cite{Jenkins_2018}. We can therefore compute all the relevant observables in terms of the parameters in our model $\lambda^\prime_i$ and $M$. Using this procedure we obtain the limits presented in Figure~\ref{fig:LFV_limits}.

\section{Di-Higgs production}

Let us discuss di-Higgs production. Considering the leading contribution in the large $\lambda^\prime$ limit, we find, following~\cite{Falkowski:2001958},
\begin{equation}
\kappa_\lambda = 1 - \frac{8 v^4 \wc_\phi}{m_H^2}+\ldots =
1 - \frac{4}{3 \pi^2} \frac{\lambda^{\prime\,6} v^4}{m_H^2 M^2}
+ \ldots
=
1- \frac{1}{6 \pi^2} \left( 2 \frac{m^{\prime\,2}}{M^2}\right)^3
\frac{M^4}{m_H^2 v^2},
\label{eq:kappa_lambda}
\end{equation}
where the dots denote other contributions that are not enhanced by the largest power of $\lambda^\prime$ and in the last equality we have written the result in terms of the mixing squared. We have taken into account that other operators that can contribute to di-Higgs production are either not generated in our model at this order or are not enhanced in the large $\lambda^\prime$ limit. The current experimental precision in $\kappa_\lambda$ is still quite poor~\cite{CMS:2025wvt}
\begin{equation}
-0.71 \leq \kappa_\lambda \leq 6.1,   \quad \text{95\% C.L..} 
\end{equation}
Nevertheless, the strong dependence on the mixing and the fact that the correction is negative, which is more strongly constrained, results in the non-trivial bound
\begin{equation}
2 \frac{m^{\prime\,2}}{M^2} \leq 
\left( 
(1 + |\kappa_{\mathrm{min}}^{\mathrm{exp}}|)
6 \pi^2 \frac{v^2 m_H^2}{M^4}
\right)^{\frac{1}{3}}
\approx \left (\frac{0.47\mbox{ TeV}}{M} \right)^{\frac{4}{3}},
\end{equation}
where we have used $\kappa_{\mathrm{min}}^{\mathrm{exp}}=-0.71$ in the second equality but we have left it general in the first one to use it for future measurements.

\bibliographystyle{JHEP}
\bibliography{main.bib}

\end{document}